\documentclass[%
 aip,
 amsmath,amssymb,
 reprint,%
]{revtex4-1}

\usepackage{graphicx}% Include figure files
\graphicspath{{figures/}}
\usepackage{dcolumn}% Align table columns on decimal point
\usepackage{bm}% bold math
\usepackage[utf8]{inputenc}
\usepackage[T1]{fontenc}
\usepackage{mathptmx}
\usepackage{etoolbox}
\DeclareMathOperator{\sgn}{sgn}

\makeatother
\begin{document}

\preprint{AIP/123-QED}

\title[]{Mode-selective Single-dipole Excitation and Controlled Routing of Guided Waves in a Multi-mode Topological Waveguide}

\author{Yandong Li}
 \thanks{Authors to whom correspondence should be addressed: Yandong Li, yl2695@cornell.edu; Gennady Shvets, gshvets@cornell.edu}
\affiliation{School of Applied and Engineering Physics, Cornell University, Ithaca, New York 14853, USA}
\author{Yang Yu}
\affiliation{School of Applied and Engineering Physics, Cornell University, Ithaca, New York 14853, USA}
\author{Kueifu Lai}
\affiliation{Department of Physics, The University of Texas at Austin, Austin, Texas 78712, USA}
\author{Yuchen Han}
\affiliation{Department of Physics, Cornell University, Ithaca NY 14853, USA}
\author{Fei Gao}
\affiliation{Division of Physics and Applied Physics, School of Physical and Mathematical Sciences, Nanyang Technological University, 21 Nanyang Link, Singapore 637371, Singapore}
\affiliation{Interdisciplinary Center for Quantum Information, State Key Laboratory of Modern Optical Instrumentation, ZJU-Hangzhou Global Science and Technology Innovation Center, College of Information Science and Electronic Engineering, Zhejiang University, Hangzhou 310027, China}

\author{Baile Zhang}
\affiliation{Division of Physics and Applied Physics, School of Physical and Mathematical Sciences, Nanyang Technological University, 21 Nanyang Link, Singapore 637371, Singapore}
\affiliation{Centre for Disruptive Photonic Technologies, The Photonics Institute, Nanyang Technological University, 50 Nanyang Avenue, Singapore 639798, Singapore}
\author{Gennady Shvets}
 \thanks{Authors to whom correspondence should be addressed: Yandong Li, yl2695@cornell.edu; Gennady Shvets, gshvets@cornell.edu}
\affiliation{School of Applied and Engineering Physics, Cornell University, Ithaca, New York 14853, USA}

\date{\today}% It is always \today, today,

\begin{abstract}
Topology-linked binary degrees of freedom of guided waves have been used to expand the channel capacity of and to ensure robust transmission through photonic waveguides. However, selectively exciting optical modes associated with the desired degree of freedom is challenging and typically requires spatially extended sources or filters. Both approaches are incompatible with the ultimate objective of developing compact mode-selective sources powered by single emitters. In addition, the implementation of highly desirable functionalities, such as controllable distribution of guided modes between multiple detectors, becomes challenging in highly compact devices due to photon loss to reflections. Here, we demonstrate that a linearly polarized dipole-like source can selectively excite a topologically robust edge mode with the desired valley degree of freedom. Reflection-free routing of valley-polarized edge modes into two spatially-separated detectors with reconfigurable splitting ratios is also presented. An optical implementation of such a source will have the potential to broaden the applications of topological photonic devices.
\end{abstract}

\maketitle

Photonic structures, such as waveguides, beam-splitters, and filters, represent an important family of optical components and devices that are crucial for compact (e.g., on-chip) generation and manipulation of light~\cite{PIC_Review:2020,PIC_Book,PIC_Review:2019}.
For optical communication applications, key characteristics of such components include bandwidth, reconfigurability, and channel capacity~\cite{communication_Review:2016}.
The latter can be enhanced by employing additional degrees of freedom (DoFs) of a light wave such as its polarization state or, in the case of a multi-mode waveguide, its mode index~\cite{Vuckovic:2021,multimode_Review:2018,Bozinovic:2013}.
Additional DoFs of a photon can be utilized for various important tasks, including creating entangled states for quantum information science applications\cite{QPIC_Review:2021,Bouchard:2021,Mittal:2021}.
However, there are numerous challenges associated with using multi-mode waveguides. For example, different modes differ by their propagation speed, resulting in temporal separation between information-carrying pulses.
In addition, preferential coupling to one specific mode often requires the emitter properties, such as the orientation of its dipole transition and its spatial location inside the waveguide, to largely match the electromagnetic profile of that mode and mismatch those of the others~\cite{Lodahl:2015,Lodahl_Review:2015,RodriguezFortuno:2013}.

Topological photonics exploits symmetries in real and reciprocal spaces to enable robust propagation of edge (or kink) modes guided by domain walls between gapped photonic crystals with different quantized topological indices~\cite{Ozawa_Review:2019,Shvets_Review:2017,Lu_Review:2014}.
Examples of such topological indices associated with propagation bands include the Chern number~\cite{Raghu:2008_1,WangZheng:2009} that can have nonzero integer values with broken time-reversal symmetry (TRS), as well as the spin-Chern~\cite{Khanikaev:2012,CTChan:2014,Ma:2015,Lai:2016,Ma:2017} and the valley-Chern~\cite{Ma:2016,GaoFei:2018,Litchinitser:2019,ZhangXiang:2017,Rechtsman:2018} half-integer indices in the systems with preserved TRS.
Topological robustness associated with the latter Chern indices is contingent on the conservation of the corresponding binary DoFs: spin and valley DoFs, respectively.
The synthetic spin DoF is preserved by the property of spin-degeneracy~\cite{Khanikaev:2012,Ma:2015}; the valley DoF is preserved by the specific orientations of domain walls or photonic crystal terminations~\cite{Ma:2016,GaoFei:2018}.

\begin{figure*}
    \centering
    \includegraphics[width=0.8\textwidth]{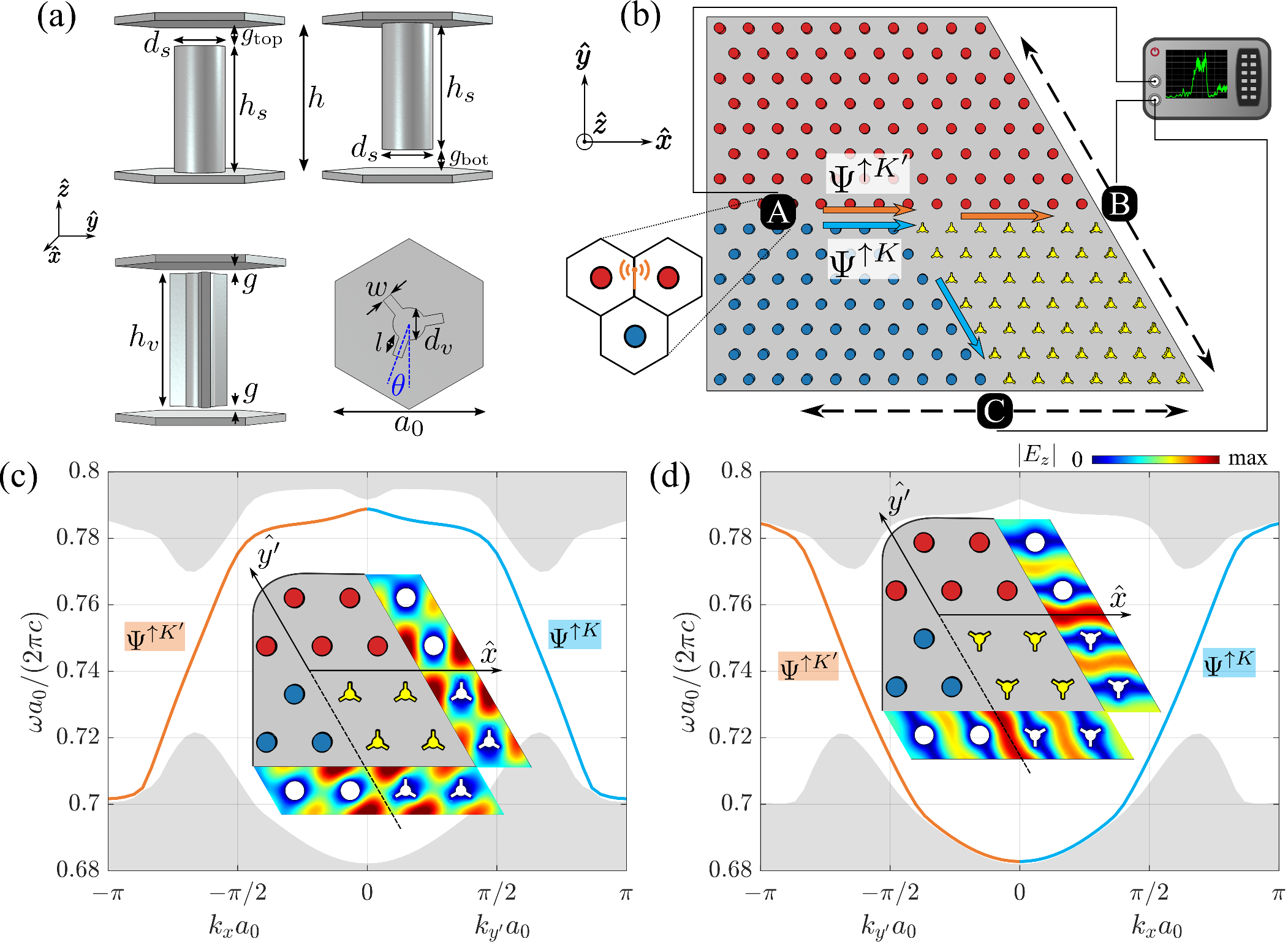}
    \caption{\label{fig:1}
    (a) Geometrical definitions of the PTI unit cells.
    Upper left (right): SPC$^{+(-)}$ with $\Delta_s>0$ ($\Delta_s<0$).
    Lower left (right): VPC side (cross section) views.
    VPC$^{+(-)}$ configuration: $\Delta_v>0$ when $\theta<30^{\circ}$ ($\Delta_v<0$ when $\theta>30^{\circ}$).
    (b) Photonic structure for valley-selective splitting of the TREK states.
    Blue (red) circles: SPC$^{+(-)}$.
    Yellow tripods: VPC with $\theta=60^{\circ}$.
    A: valley-selective emitter embedded inside the MMW.
    B and C: receiving ports.
    Orange (cyan) arrow: $\Psi^{\uparrow K'}$ ($\Psi^{\uparrow K}$) TREK states.
    (c) Band diagrams of the two SPC-VPC interfaces from (b), for the tripods configuration $\theta=60^{\circ}$.
    Only the spin up states excited by emitter A in the SPC$^+$-SPC$^-$ MMW are shown.  Orange (cyan) solid lines: band-gap-spanning TREK states with $v=K'$ ($v=K$) valley DoF.
    Gray background: bulk modes.
    (d) Same as (c), but for the tripods configuration $\theta=0^{\circ}$.
    In (c,d), half of the band diagram is for $k_{y'}$ which is along the $y'$-direction labeled in insets.
    Insets: $|E_z|$ distributions of the TREK states at the corresponding SPC-VPC interfaces.
    Unit cell parameters: $d_s=0.345a_0$, $h_s=0.85a_0$, $g_{\text{top}} = g_{\text{bot}} = 0.15a_0$, $l=0.12a_0$, $w=0.06a_0$, $d_v=0.2a_0$, $g=0.03a_0$, $h_v=0.94a_0$, $h=a_0$, where $a_0=36.8$mm is the lattice constant.
    }
\end{figure*}

Therefore, topological photonics provides an entirely new way of thinking about propagation robustness, reflection suppression, and other empirically useful phenomena that can be engineered through preserving binary DoFs.
For example, a domain wall between two photonic crystals with different topological indices can support topologically robust edge or kink (TREK) states. These TREK states are associated with conserved DoFs and do not suffer from back-scattering. Not surprisingly, advances in topological photonics have already contributed to conceptual developments in numerous photonic components and devices~\cite{Anlage_Review:2020}, including waveguides~\cite{Litchinitser:2019}, cavities~\cite{PengChao:2019,Ota:19,LuLing:2019,Yandong:2020}, and lasers~\cite{TIlaser:Theory,TIlaser:Experiment,ZhangBaile:2020}.
Here we utilize the ideas from topological photonics to develop an approach to mode-selective excitation of multi-mode robust waveguides by single (point-like) emitters.
We also demonstrate reflection-free routing (or beam-splitting) of the excited modes into two spatially-separated detectors [see schematic in Fig.~\ref{fig:1}(b)] with reconfigurable splitting ratios.
Unbalanced beam splitting has been proposed and used in several multi-photon non-classical interference experiments~\cite{Cerf:2020,Zeilinger:2006}.

The multi-mode waveguide (MMW) is formed by the domain wall between two spin photonic crystals (SPCs) with opposite signed spin Chern number $C_{\rm SPC}^{s,v}$, where $s=\uparrow,\downarrow$ labels the binary spin DoF, and $v=K, K^{\prime}$ labels the binary valley DoF.
This MMW supports two forward-propagating ($s=\uparrow$) TREK states with identical group velocities, but different phase velocities corresponding to $v=K$ and $v=K^{\prime}$~\cite{Ma:2015,Lai:2016,Ma:2017}.
The routing capability is enabled by placing a rhomboid-shaped valley photonic crystal (VPC) at the corner of the SPC domains.
Domain walls between the band-gap-sharing SPC and VPC regions serve as single-mode waveguides supporting chiral TREK states labeled by their conserved spin and valley DoFs.

As a specific platform for realizing the SPCs and VPCs, we adopt the microwave photonic crystals introduced in Ref.~\cite{Ma:2017} and used in several experiments~\cite{Lai:2016,GaoFei:2018}.
The structure is comprised of an array of metallic rods arranged in a triangular lattice and placed between two metallic plates (at $z=\pm h/2$) which confines wave along the $\hat{z}$-direction.
Non-trivial spin and valley textures are produced by breaking two distinct spatial symmetries~\cite{Ma:2017}.
Specifically, by placing the cylindrical rods asymmetrically with respect to the mid-height ($z=0$) plane, the mirror symmetry is broken, and nonzero valley-independent photonic spin-Chern indices $C_{\rm SPC}^{\uparrow (\downarrow),v} = \pm \sgn(\Delta_s)/2$ are induced.
Here the $+$ ($-$) signs correspond to $\uparrow$ ($\downarrow$) spin-polarized states and $\sgn(\Delta_s) = \sgn(g_{\text{top}} - g_{\text{bot}})$ [Fig.~\ref{fig:1}(a)], and $|\Delta_s| \propto (h-h_s)$ is proportional to the SPC band gap width (see Supplementary Material).
Likewise, by modifying the rods' cross sections from circles to $C_{3v}$-symmetric tripod-like shapes, the in-plane inversion symmetry is broken, and nonzero spin-independent valley-Chern indices $C_{\rm VPC}^{s,K(K')} = \pm \sgn(\Delta_v)/2$ are induced. The $+$ ($-$) signs correspond to $K (K^{\prime})$ valley-polarized states. $|\Delta_v| \propto (\theta - 30^{\circ})$ [Fig.~\ref{fig:1}(a)] corresponds to the VPC band gap width (see Supplementary Material).

The bulk-edge correspondence predicts that, at SPC-VPC interfaces, only the TREK states with specific combinations of spin- and valley-DoFs exist~\cite{Ma:2017,KangYuhao:2018,GaoFei:2018}.
As an example, we consider the SPC$^-$-VPC$^-$ interface [along the $\hat{x}$-direction in Fig.~\ref{fig:1}(c)].
Here the $-$ sign represents $\Delta_s < 0$ ($\Delta_v < 0$) for SPC (VPC), respectively.
For the subspace of electromagnetic modes with $s=\uparrow$ and $v=K'$, the difference between the Chern indices is $\Delta C = |C_{\rm SPC}^{\uparrow, K'} - C_{\rm VPC}^{\uparrow, K'}| = 1$, indicating that this interface supports one forward-propagating TREK state marked as $\Psi^{\uparrow K'}$ in Fig.~\ref{fig:1}(c), where $+\hat{x}$ is the forward direction.
Similarly, the SPC$^+$-VPC$^-$ interface supports a $\Psi^{\uparrow K}$ TREK state propagating along the $-\hat{y'} = \hat{x}/2 - \sqrt{3}\hat{y}/2$ direction [Fig.~\ref{fig:1}(c)].

Propagation directions of such TREK states are exchanged by flipping the sign of $\Delta_v$ of VPC: the $\Psi^{\uparrow K}$ ($\Psi^{\uparrow K'}$) state propagates along the $+\hat{x}$($-\hat{y'}$)-direction [Fig.~\ref{fig:1}(d)].
For the specific VPC geometry shown in Fig.~\ref{fig:1}(a), such sign flip is accomplished by rotating the tripods.
Therefore, the orientation angle $\theta$ of the tripods inside the VPC domain [Fig.~\ref{fig:1}(b)] can be used for valley-dependent splitting of electromagnetic energy: depending on $\theta$, a given valley component predominantly flows into either port B or C.

Valley-dependent beam splitting using large VPC domains prevents the TREK states from evanescent tunneling into an ``unintended'' port, and produces essentially $\sim 100\%$ flow into an ``intended'' port~\cite{KangYuhao:2018}.
In contrast, when the VPC domain is a compact, ``poor" insulator (with a narrow band gap), the tunneling effect becomes significant.
The finite size of the VPC region enables tunneling of valley-polarized TREK states into the unintended port, thereby allowing the energy flow ratio to vary in a wide range.
Using semi-analytical methods, we find that such change results from the considerable evanescent tunneling when the band gap is narrow.

TREK states can be excited by feeding optical energy into the structure either from an external source~\cite{Litchinitser:2019} (e.g., through a waveguide) or directly by dipole-like emitters~\cite{GaoFei:2018} embedded inside the topological MMW.
The latter approach can potentially lead to more compact devices.
However, exciting a spin-valley-polarized TREK state $\Psi^{s v}$ with a specific combination of spin and valley DoFs $(s,v)$ by a single emitter can be challenging.
An arbitrarily placed point source (e.g., a single quantum emitter for the optical, or an electrically-small antenna for the microwave frequency range) will, in general, couple to multiple modes supported by the MMW.
For example, a $\hat{z}$-polarized dipole placed inside a SPC$^+$-SPC$^-$ MMW generally radiates into all of the four TREK modes $\Psi^{s v}$ modes, where $s=\uparrow,\downarrow$ and $v=K,K^{\prime}$.
For the TREK states excited by source A and traveling towards the beam splitter [indicated as the two arrows in Fig.~\ref{fig:1}(b)], the spin-DoF must be $s=\uparrow$.
Although the sign of the group velocity automatically selects the spin-DoF, the valley-DoF is still undetermined.

While it is possible to block one of the two valley-polarizations by adding an extra filter to the structure~\cite{KangYuhao:2018}, such an approach compromises compactness and completely prevents one of the two valley-polarized photons ($v=K$ or $v=K^{\prime}$) from entering the beam splitter.
An alternative approach to exciting the TREK state with a specific valley polarization utilizes precisely phased radiation sources~\cite{Anlage:2016,GaoFei:2018}.
However, this approach can hardly be compatible with the goal of developing single emitters for on-chip optical and potentially quantum information processing applications.

\begin{figure}
    \centering
    \includegraphics[width=0.48\textwidth]{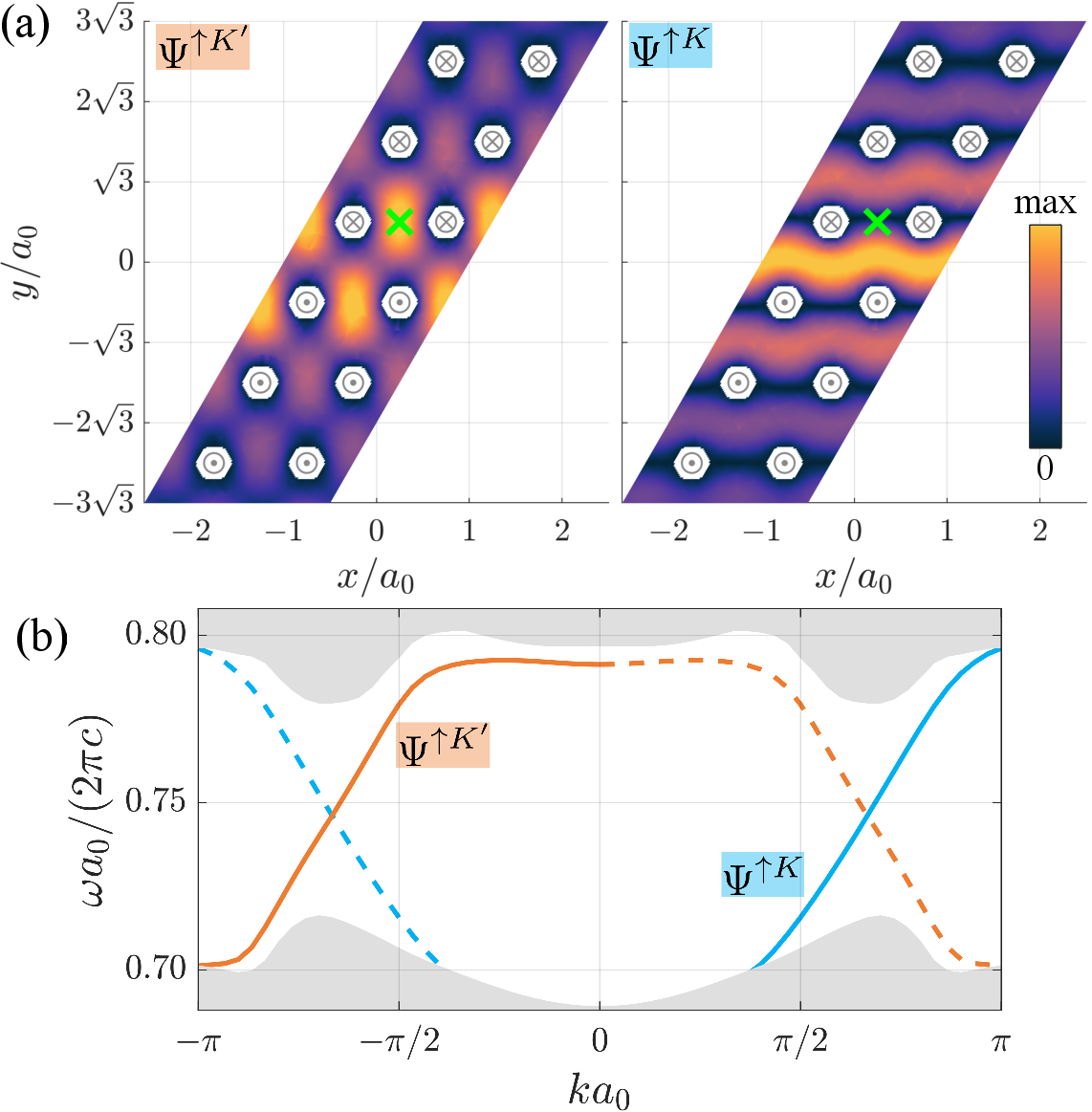}
    \caption{\label{fig:2} TREK mode profiles and the photonic band structure of the SPC$^+$-SPC$^-$ MMW.
    (a) $|E_z|$ distribution of the $\Psi^{\uparrow K'}$ (left) and $\Psi^{\uparrow K}$ (right) TREK states in the mid-height ($z=0$) plane at $\omega = \omega_0$.
    Green crosses: source position $\mathbf{r}_{K'}$ for selectively exciting the $K'$-valley-polarized state.
    Symbol $\bigotimes$ ($\bigodot$): rods attached to the top (bottom) plate.
    (b) Photonic band structure: bulk (gray shading) and TREK (solid and dashed lines) states.
    Solid orange (cyan) lines: forward-propagating $\Psi^{\uparrow K'}$ ($\Psi^{\uparrow K}$) states.
    Dashed lines: backward-propagating states moving away from the beam splitter.
    Parameters: same as in Fig.~\ref{fig:1}.
    Mid-gap frequency: $\omega_0 = 0.744 \left( 2\pi c/a_0 \right)$.}
\end{figure}

Here we demonstrate that a linearly polarized dipole-like source can selectively excite valley-polarized TREK states propagating from the emitter towards the valley-dependent beam splitter.
The latter can distribute such polarized states between two output ports in a controllable proportion.
We place the emitter at a judiciously designed position based on the electromagnetic field distribution of the desired valley-polarized state.
The $\hat{z}$-directional electric field distributions of the $\Psi^{\uparrow K'}$ and $\Psi^{\uparrow K}$ states are profoundly different [Fig.~\ref{fig:2}(a)], thus enabling valley-selective excitation of TREK states using a $\hat{z}$-directional electric dipole.

By placing the dipole source oscillating at frequency $\omega$ inside the band gap at the position $\mathbf{r}$ such that $|E_z^{\uparrow K'}(\omega,\mathbf{r})| \gg |E_z^{\uparrow K}(\omega,\mathbf{r})|$, we expect the excitation efficiency $\eta^{\uparrow K'}(\omega) \propto |E_z^{\uparrow K'}(\omega,\mathbf{r})|^2$ of the $\Psi^{\uparrow K'}$ state to be much larger than that of $\Psi^{\uparrow K}$ (see Supplementary Material for details).
Specifically, when the dipole source is placed at the position $\mathbf{r} = \mathbf{r}_{K^{\prime}}$ [Fig.~\ref{fig:2}(a)], we find that $\eta^{\uparrow K'}/\eta^{\uparrow K} \approx 43$ at the mid-gap frequency $\omega = \omega_0 = 0.744 \left( 2\pi c/a_0 \right)$.
The high degree of valley selectivity can be verified by detecting the TREK states at the output ports B and C.

All the metallic components for the photonic platform shown in Fig.~\ref{fig:1} -- round rods for the SPC, tripod-shaped rods for the VPC, and two plates --  were constructed from aluminum.
For the SPC, the round rods were directly attached to one of the two metal plates, so either one of the rod-plate gaps was set to zero ($g_{\rm top}=0$ or $g_{\rm bot}=0$).
The tripods were symmetrically separated from both plates by foam (see Methods for details).
The operating frequency $f\equiv \omega_0/2\pi \approx 6.1 {\rm GHz}$ was set to the middle of the spectral band gap shared by SPC and VPC.

First, we experimentally validated the concept of valley-selective TREK launching at the SPC$^{+}$-SPC$^{-}$ domain wall [Fig.~\ref{fig:1}(b)].
This was done by placing a $z$-directional antenna between two rods in the SPC$^{-}$ domain adjacent to the SPC$^{+}$-SPC$^{-}$ domain wall, as shown in Fig.~\ref{fig:2}(a).
Tunneling through the VPC region was suppressed by selecting the tripods orientations, $\theta \equiv \theta_1 = 0^{\circ}$ and $\theta \equiv \theta_2 = 60^{\circ}$ to ensure the widest VPC band gaps.
When the tunneling is minimized, only one port (B for $\theta = \theta_2$ and C for $\theta = \theta_1$) is expected to receive the overwhelming majority of the $K^{\prime}$-polarized radiation as illustrated by Figs.~\ref{fig:1}(c,d).

\begin{figure}
    \centering
    \includegraphics[width=0.48\textwidth]{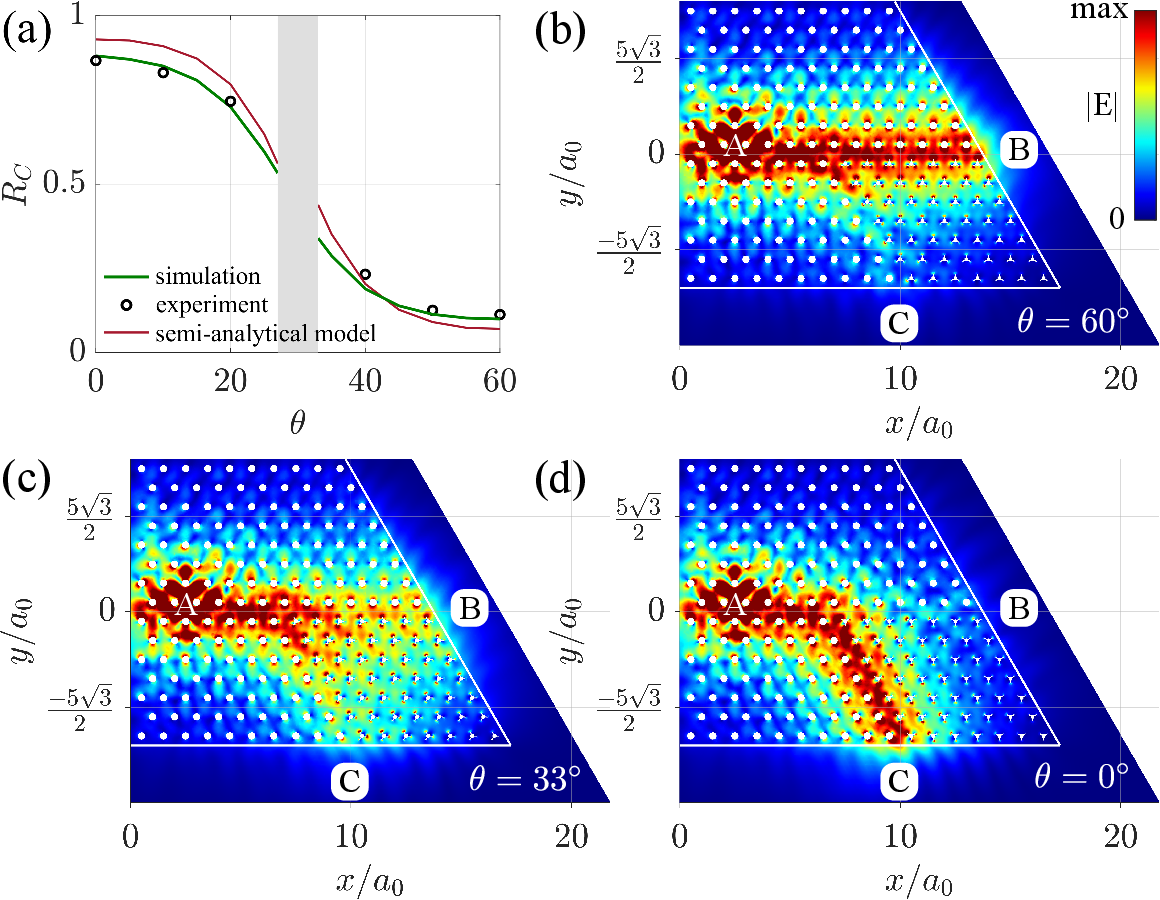}
    \caption{\label{fig:3}
    (a) Beam-splitting ratio $R_C \equiv W_C /\left( W_{C} + W_{B} \right)$: experiment (black circles), COMSOL simulation (green lines), and semi-analytical theory (red lines).
    The VPC domain has no complete band gap for $27^{\circ} < \theta < 33^{\circ}$ (shadowed region).
    (b,d) Time-averaged energy density for the largest VPC band gap.
    Near-complete transmission of the $\Psi^{\uparrow K'}$ state to port B for $\theta=60^{\circ}$ ($\Delta_v < 0$) and to port C for $\theta=0^{\circ}$ ($\Delta_v > 0$).
    (c) Same as (b), but for a narrow VPC band gap for $\theta=33^{\circ}$.
    $R_C^{\rm sim} \approx 0.34$.
    White lines: periphery of the metallic structure.
    The domain outside of the structure is free space.
    }
\end{figure}

The tunneling effect is quantified by introducing the power fluxes $W_{B(C)}$ captured by the corresponding port.
The experimentally measured quantities $W_{B(C)}^{\rm exp}$ are obtained by scanning the pick-up antenna attached to the moving stage as shown in Fig.~S7 of the Supplementary Material.
The experimental beam-splitting ratio, defined as $R_C^{\rm exp} \equiv W_C^{\rm exp}/\left( W_C^{\rm exp} + W_B^{\rm exp} \right)$, is shown as circles in Fig.~\ref{fig:3} for a range of tripod configuration angles.
Specifically, $R_C^{\rm exp} \approx 0.87$ when $\theta=\theta_1$, and $R_B^{\rm exp} \equiv 1-R_C^{\rm exp} \approx 0.89$ when $\theta=\theta_2$, confirming that the experimentally excited TREK state is indeed predominantly $\Psi^{\uparrow K'}$.
We note that a backward-propagating $\Psi^{\downarrow K}$ TREK state of equal magnitude is also excited, because $\Psi^{\uparrow K'}$ and $\Psi^{\downarrow K}$ are time-reversal conjugates~\cite{Ma:2017,Zayats:2019}.
However, $\Psi^{\downarrow K(K')}$ propagates away from the VPC and cannot be measured by the detectors at port B or C.

For comparison, we simulated the structure with COMSOL Multiphysics and calculated the corresponding quantities $R_C^{\rm sim} \equiv W_C^{\rm sim}/\left( W_C^{\rm sim} + W_B^{\rm sim} \right)$ [Fig.~\ref{fig:3}(a)].
$W_{B(C)}^{\rm sim}$ is defined as $W_{B(C)}^{\rm sim} \equiv \iint_{B(C)} dS \langle W \rangle$, The surface integral is over the structural periphery at the corresponding port, and $\langle W \rangle$ is the time-averaged power flow perpendicular to that periphery.
Simulation results are in good agreement with the experiment: $R_C^{\rm sim} \approx 0.88$ when $\theta=\theta_1$, and $R_B^{\rm sim} \equiv 1-R_C^{\rm sim} \approx 0.90$ when $\theta=\theta_2$.
Note that a balanced beam splitter corresponds to $R_C = R_B = 0.5$, i.e., it splits energy of the incoming TREK state equally between the two ports.

An important conclusion from these findings is that, even though the tunneling of the TREK states propagating towards their ``intended'' port into the ``unintended'' port is minimized by maximizing the VPC band gap for $\theta = \theta_{1,2}$, it cannot be entirely eliminated for a relatively compact beam-splitter used in this work.
The tunneling causes an effective ``averaging'' effect, lowering the received energy ratio $R_{C(B)}$ [for $\theta=\theta_{1(2)}$] below the excitation efficiency ratio $\eta^{\uparrow K'}/\left( \eta^{\uparrow K'} + \eta^{\uparrow K} \right) \approx 0.98$.
This is because, with the maximized VPC band gap, the tunneling is still not negligible in this finite-sized structure (see Supplementary Material for details).

As shown in Fig.~\ref{fig:3}, this tunneling effect becomes more significant when the band gap narrows for $\theta_1 < \theta \leq \theta_g$ or $\theta_g \leq \theta < \theta_2$, where $\theta_g = 30^{\circ}$ corresponds to the tripods orientation that closes the band gap.
The experimentally measured and simulated beam splitting ratios change smoothly as $\theta$ approaches $\theta_g$, and the band gap approaches zero.
For example, when $\theta = 33^{\circ}$, the exponential decay of the $\Psi^{\uparrow K'}$ state along the $-\hat{y}$-direction becomes significant (Fig.~\ref{fig:3}c), causing considerable amount of energy to tunnel to Port C.
Below we develop a semi-analytical theory that describes how electromagnetic energy is distributed between the two ports in the presence of tunneling.

\begin{figure}
    \centering
    \includegraphics[width=0.48\textwidth]{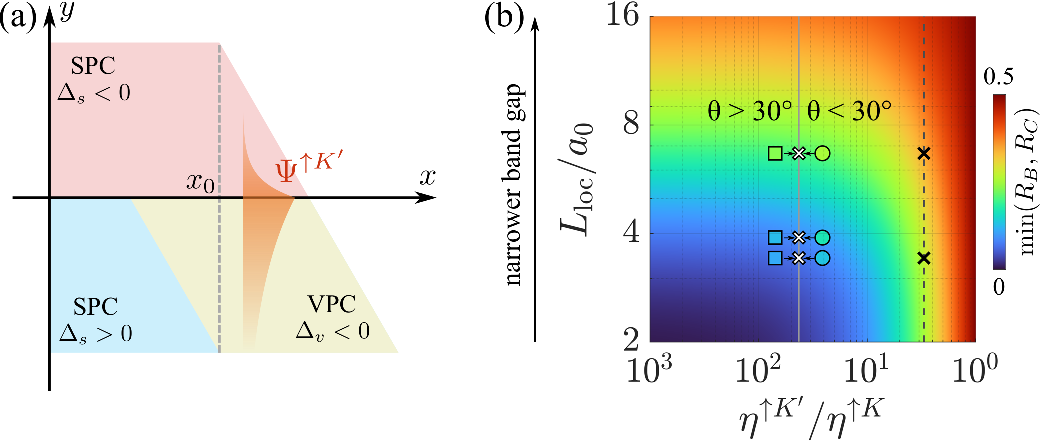}
    \caption{\label{fig:4}
    (a) Schematic: leakage of the $\Psi^{\uparrow K'}$ state through the finite-sized VPC (yellow) region. Red and blue regions: SPCs with opposite signed spin-Chern numbers.
    (b) The beam-splitting ratio (color-coded) of the minor port [port $B$($C$) for $0^{\circ}<\theta<30^{\circ}$ ($30^{\circ}<\theta<60^{\circ}$)].
    High (low) contrast between the two ports occurs when $\min(R_B,R_C) \rightarrow 0$ [$\min(R_B,R_C) \rightarrow 0.5$].
    Horizontal axis: valley-polarization selectivity of the excitation, $\eta^{\uparrow K'}/\eta^{\uparrow K}$.
    Vertical axis: localization distance $L_{\rm loc}$ of the TREK state inside the VPC, determined by the VPC band gap.
    Filled circles (squares): filling-color-coded experimental data of $R_{B(C)}$ for $\theta = 0^{\circ}$, $10^{\circ}$, and $20^{\circ}$ ($\theta = 40^{\circ}$, $50^{\circ}$, and $60^{\circ}$), plotted to the right (left) of the gray vertical line which represents $\eta^{\uparrow K'}/\eta^{\uparrow K} \approx 43$, corresponding to the source shown in Fig.~\ref{fig:2}.
    The dashed black vertical line represents $\eta^{\uparrow K'}/\eta^{\uparrow K} = 3$ for comparison.
    The two black crosses label the maximum and minimum localization distances ($L_{\rm loc,max}=6.7a_0$, $L_{\rm loc,min}=3.4a_0$) that we implemented in experiment.}
\end{figure}

To quantitatively describe the tunneling phenomena, we use the following ansatz to represent the $\Psi^{\uparrow K'}$ TREK state propagating along the SPC-VPC domain wall towards detector B:
\begin{equation}
\label{eq:ansatz}
\Psi^{\uparrow K'}(x,y,t)=
\begin{cases}
A e^{-i\omega t + (ik - \kappa_x) x} e^{-\kappa_2 y}, & \text{in the SPC domain}, \\
A e^{-i\omega t + (ik - \kappa_x) x} e^{\kappa_1 y}, & \text{in the VPC domain},
\end{cases}
\end{equation}
for $x > x_0$ as defined in Fig.~\ref{fig:4}a.
Here $\omega$ and $k$ are the angular frequency and the wave-vector, respectively, and $\kappa_{1(2)}$ is the decay constant into the VPC (SPC) domain along the transverse ($\pm \hat{y}$) direction.
The resulting tunneling towards the structural periphery naturally leads to energy loss.
Therefore, the amplitude of the TREK state also exponentially decays along its propagation $\hat{x}$-direction with the decay constant $\kappa_x$.
Throughout this study, $\kappa_2 \equiv \kappa_2(\Delta_s)$ is a constant determined by the band gap width of the SPC.
On the other hand, $\kappa_1 \equiv L_{\rm loc}^{-1}(\Delta_v)$ implicitly depend on $\theta$, where $L_{\rm loc}$ is the localization distance along the transverse direction (see Supplementary Material for details).
As the VPC band gap narrows, $L_{\rm loc}$ increases.

Because $\kappa_x$ and $\kappa_1$ are physically related to each other -- finite $\kappa_x$ is the consequence of the transverse directional tunneling that happens when the wave propagates in the finite-sized structure along the longitudinal direction -- we use a semi-empirical equation to relate $\kappa_x$ and $\kappa_1$: $\kappa_x = \kappa_{x0} e^{-t \kappa_1}$ (see Supplementary Material for details).
Here $\kappa_{x0} \approx 0.12/a_0$ and $t \approx 7.9 a_0$.
The energy measured at port B (C), $W_{B(C)}$, is a sum of the \textit{transmitted} energy $W_{\rm tr}$ and the \textit{tunneled} energy $W_{\rm tu}$:
\begin{equation}
\begin{aligned}\label{eq:energy}
& W_C = \eta^{\uparrow K'} W_{\rm tr} + \eta^{\uparrow K} W_{\rm tu},\ \text{and}, \\
& W_B = \eta^{\uparrow K} W_{\rm tr} + \eta^{\uparrow K'} W_{\rm tu}, & \text{for}\ 0^{\circ}<\theta<30^{\circ}; \\
& W_C = \eta^{\uparrow K} W_{tr} + \eta^{\uparrow K'} W_{\rm tu},\ \text{and}, \\
& W_B = \eta^{\uparrow K'} W_{\rm tr} + \eta^{\uparrow K} W_{\rm tu}, & \text{for}\ 30^{\circ}<\theta<60^{\circ},
\end{aligned}
\end{equation}
where $W_{\rm tr}$ and $W_{\rm tu}$ depend on the decay constants $\kappa_1$, $\kappa_2$, and $\kappa_x$.
Here $\eta^{\uparrow K'(K)}$ is the excitation efficiency of the $\Psi^{\uparrow K'(K)}$ state in the SPC$^+$-SPC$^-$ MMW.

With Eq.~\ref{eq:energy} and the decay constants calculated from the numerical results of COMSOL simulation (see Supplementary Material for details), we compute the ratio of energy that port C receives, $R_C$.
The result [red curve in Fig.~\ref{fig:3}(a)] reveals the ``averaging'' effect due to band gap narrowing: the difference between $W_B$ and $W_C$ becomes smaller when $\theta$ approaches $30^{\circ}$.
We note that the tripod configuration angles $\theta$ and $60^{\circ} - \theta$ correspond to the same VPC band gap width and hence the same $L_{\rm loc}$.
Therefore, the semi-analytical result satisfies $R_B(\theta) = R_C(60^{\circ}-\theta)$.
We focus on the energy ratio of the minor port [the one receives less energy, also see Fig.~\ref{fig:4}(b) caption], $\min(R_B,R_C)$.

Figure~\ref{fig:4}(b) demonstrates how $\min(R_B,R_C)$ depends on the excitation efficiency ratio $\eta^{\uparrow K'} / \eta^{\uparrow K}$ and the localization distance $L_{\rm loc}$ inside the VPC.
By comparing the experimental results (filled dots) with the semi-analytical model (colored background along the gray solid line), we observe that our model accurately captures how $\min(R_B,R_C)$ depends the localization length.
Furthermore, Fig.~\ref{fig:4}(b) reveals that the high excitation selectivity grants considerable tunability of $R_{B(C)}$ for this compact structure.

The judiciously positioned dipole source provides an excitation efficiency ratio as high as $\eta^{\uparrow K'} / \eta^{\uparrow K} \approx 43$.
With the high excitation selectivity of the $\Psi^{\uparrow K'}$ state, our semi-analytical model predicts that $\min(R_B,R_C)$ can change from $0.07$ for $L_{\rm loc} = 3.4a_0$ (insignificant tunneling, high contrast between the two ports) to $0.20$ for $L_{\rm loc} = 6.7a_0$ (significant tunneling, moderate contrast between the two ports).
In experiment, as $\theta$ changes from $60^{\circ}$ (corresponding to $L_{\rm loc} = 3.4a_0$) to $40^{\circ}$ (corresponding to $L_{\rm loc} = 6.7a_0$), the energy ratio of the minor port is tuned from $R_C \approx 0.11$ to $R_C \approx 0.23$; as $\theta$ changes from $0^{\circ}$ to $20^{\circ}$, this ratio is tuned from $R_B \approx 0.13$ to $R_B \approx 0.25$.
The result demonstrates that, for this compact structure, the energy received by the minor port can increase by $\sim 100\%$ as $L_{\rm loc}$ doubles.
On the other hand, if the excitation selectivity was merely $\eta^{\uparrow K'}/\eta^{\uparrow K} \approx 3$, then, based on the semi-analytical result, $\min(R_B,R_C)$ can only be tuned from $0.27$ to $0.34$, i.e., only by $\sim 26\%$.
As demonstrated in Fig.~\ref{fig:4}(b), the color contrast along the dashed black line ($\eta^{\uparrow K'}/\eta^{\uparrow K} \approx 3$) is much less significant than that along the 
gray line ($\eta^{\uparrow K'}/\eta^{\uparrow K} \approx 43$).
Overall, the high selectivity of the excitation enables the remarkable tunability of $R_{B(C)}$ for this compact structure.

In summary, we demonstrate an approach for manipulating the edge mode
supported by a topological MMW.
The waveguide is a sharp domain wall between two SPCs with opposite spin-Chern numbers, which supports two TREK states co-propagating with identical group velocities and distinguished by their binary valley DoFs.
In addition to the topological robustness of the TREK states, we have demonstrated another important feature of this MMW:
a single linearly polarized emitter can excite, with high selectivity, only one of the two co-propagating TREK states.
The high selectivity of excitation is verified by feeding the excited states into a compact junction of two topologically different SPC-VPC interfaces directed into separate output ports.

By controlling the band gap width of the VPC domain of the valley-DoF-sorter, it is possible to re-distribute the energy of a valley-polarized TREK state between the two output ports over a wide range: from $\sim 1:8$ to $\sim 1:3$, corresponding to the widest and the narrowest experimentally realized band gaps, respectively.
We have presented a concise semi-analytical model to demonstrate that the physical reason for the energy re-distribution between the two output ports is evanescent tunneling.
The model could pave the way for applications in optical, valleytronic information processing (see Supplementary Material for details).
We expect that future efforts to extend the concept of valley-based MMWs with single emitter source to the optical spectrum will benefit various applications from telecommunications to quantum information science.

\section*{Supplementary Material}
See Supplementary Material for details about the semi-analytical model, calculation of the excitation efficiency, the experimental setup, and future prospects for applications in optical valleytronics.

\begin{acknowledgments}
This work was supported by the Office of Naval Research (ONR) Award No. N00014-21-1-2056 and the Army Research Office (ARO) Award W911NF2110180. Helpful discussions with Minwoo Jung and Maxim Shcherbakov are gratefully acknowledged.
\end{acknowledgments}

\section*{Data Availability Statement}
The data that support the findings of this study are available from the corresponding author upon reasonable request.

%

%%%%%%%%%% Merge with supplemental materials %%%%%%%%%%
\pagebreak
\widetext
\begin{center}
\textbf{\large Supplementary Material for\\ ``Mode-selective Single-dipole Excitation and Controlled Routing
of Guided Waves in a Multi-mode Topological Waveguide''}
\end{center}
%%%%%%%%%% Merge with supplemental materials %%%%%%%%%%
%%%%%%%%%% Prefix a "S" to all equations, figures, tables and reset the counter %%%%%%%%%%
\setcounter{equation}{0}
\setcounter{figure}{0}
\setcounter{table}{0}
\setcounter{page}{1}
\makeatletter
\renewcommand{\theequation}{S\arabic{equation}}
\renewcommand{\thefigure}{S\arabic{figure}}
%\renewcommand{\bibnumfmt}[1]{[S#1]}
%\renewcommand{\citenumfont}[1]{S#1}
%%%%%%%%%% Prefix a "S" to all equations, figures, tables and reset the counter %%%%%%%%%%

\section*{Analytical description of the edge mode at the QVH-QSH PTI interface}
The bulks of the valley photonic crystal (VPC) and the spin photonic crystal (SPC) are described by the following $8 \times 8$ Kane-Mele Hamiltonian with different perturbations~\cite{Kane_Mele:05,MaTzuhsuan:15,MaTzuhsuan:17},
\begin{equation}
\label{eq:Hamiltonian_1}
H = v (\delta k_x \hat{\tau}_z \otimes \hat{s}_0 \otimes \hat{\sigma}_x + \delta k_y \hat{\tau}_0 \otimes \hat{s}_0 \otimes \hat{\sigma}_y) + \Delta H,
\end{equation}
where $\delta \boldsymbol{k} = \boldsymbol{k} - \boldsymbol{K}$ or $\boldsymbol{k} - \boldsymbol{K}'$.
$\hat{\tau}_i$, $\hat{s}_i$, and $\hat{\sigma}_i$ ($i=0,x,y,z$) are Pauli matrices acting on the band, spin, and valley subspaces.
$v$ is the slope of the ungapped Dirac cone.
The overall basis of this bulk Hamiltonian is $[\psi^{R, \uparrow}_K, \psi^{L, \uparrow}_K, \psi^{R, \downarrow}_K, \psi^{L, \downarrow}_K, \psi^{L, \uparrow}_{K'}, \psi^{R, \uparrow}_{K'}, \psi^{L, \downarrow}_{K'}, \psi^{R, \downarrow}_{K'}]^T$, where $R(L)$ represents right (left) circular polarization.
The perturbation term is $\Delta H = H_s \propto \Delta_s \hat{\tau}_z \otimes \hat{s}_z \otimes \hat{\sigma}_z$ for SPC and $\Delta H = H_v \propto \Delta_v \hat{\tau}_0 \otimes \hat{s}_0 \otimes \hat{\sigma}_z$ for VPC.

Notice that the two differently perturbed Hamiltonians are both block-diagonal, i.e., they can be written as,
\begin{equation}
\begin{pmatrix}\label{eq:Hamiltonian_2}
H^{\uparrow K} & & &\\
& H^{\downarrow K} & &\\
& & H^{\uparrow K'} &\\
& & & H^{\downarrow K'}
\end{pmatrix},
\end{equation}
where the empty entries are $\mathbf{0}_{2 \times 2}$.

The spin-valley conservation of the edge mode manifests itself in the simultaneous block-diagonalization of the two bulk Hamiltonians.
First, we consider the SPC$^-$-VPC$^-$ interface ($\Delta_s < 0$ and $\Delta_v < 0$).
In this scenario, only the $2$nd and $3$rd diagonal entries ($H^{\downarrow}_K$ and $H^{\uparrow}_{K'}$) support edge modes, indicating that this type of interface only supports $\Psi^{\uparrow K'}$ and $\Psi^{\downarrow K}$ edge modes.
For example, we focus on the valley $K'$ and spin $\uparrow$ polarization.
The two PTIs are described by the following Hamiltonians,
\begin{equation}
\begin{aligned}\label{eq:Hamiltonian_3}
H^{\uparrow K'}_{QVH} \equiv H_1 = v (-\delta k_x \hat{\sigma}_x + \delta k_y \hat{\sigma}_y) - m_1 \hat{\sigma}_z, \\
H^{\uparrow K'}_{QSH} \equiv H_2 = v (-\delta k_x \hat{\sigma}_x + \delta k_y \hat{\sigma}_y) + m_2 \hat{\sigma}_z.
\end{aligned}
\end{equation}

Consider the envelope-function equation for the interface~\cite{TIbook:2015},
\begin{equation}
\begin{aligned}\label{eq:EFA}
& [- \hat{p}_x \hat{\sigma}_x + \hat{p}_y \hat{\sigma}_y + m(y) \hat{\sigma}_z] \phi(x,y) = E\phi(x,y), \\
& m(y)=
\begin{cases}
m_2, & \text{if}\ y>0, \\
-m_1, & \text{if}\ y<0,
\end{cases}
\end{aligned}
\end{equation}
where $\hat{p}_x = -i\partial / \partial x$, $\hat{p}_y = -i\partial / \partial y$, $m_2 > 0$, and $m_1 > 0$.

The wavefunction ansatz is,
\begin{equation}
\label{eq:wavefunction}
\phi(x,y) = t
\begin{pmatrix}
a_1\\
b_1
\end{pmatrix} e^{i k_x x} e^{\kappa_1 y} \Theta(-y) + 
\begin{pmatrix}
a_2\\
b_2
\end{pmatrix} e^{i k_x x} e^{-\kappa_2 y} \Theta(y),
\end{equation}
where $\Theta$ represents the Heaviside step function and $t$ is an undetermined parameter.
$(a_1,b_1)^T$ and $(a_2,b_2)^T$ are the eigenvectors of $H^{\uparrow K'}_{QVH}$ and $H^{\uparrow K'}_{QSH}$.
For the positive energy solution,
\begin{equation}
\begin{pmatrix}
a_1\\
b_1
\end{pmatrix}
= \begin{pmatrix}
m_1 - E\\
v(k_x - \kappa_1)
\end{pmatrix}
\quad\text{and}\quad 
\begin{pmatrix}
a_2\\
b_2
\end{pmatrix}
= \begin{pmatrix}
- m_2 - E\\
v(k_x + \kappa_2)
\end{pmatrix}
\end{equation}

Solving Eq.~\ref{eq:EFA} with the ansatz in Eq.~\ref{eq:wavefunction}, we find,
\begin{equation}
\begin{aligned}\label{eq:step_1}
& E = \sqrt{m_1^2 + v^2(k_x^2 - \kappa_1^2)} = \sqrt{m_2^2 + v^2(k_x^2 - \kappa_2^2)}, \\
& \frac{m_1 - E}{-m_2 - E} = \frac{k_x - \kappa_1}{k_x + \kappa_2}.
\end{aligned}
\end{equation}

Although Eq.~\ref{eq:step_1} has four different solutions,
\begin{equation}
\begin{aligned}\label{eq:step_2}
& \kappa_1 = \pm m_1/v,\ \kappa_2 = \pm m_2/v,\ E = \pm k_x v, \\
& \kappa_1 = k_x,\ \kappa_2 = \pm \sqrt{-m_1^2 + m_2^2 + k_x^2v^2}/v,\ E=m_1,
\end{aligned}
\end{equation}
the physical solution of the edge mode should have energy within the band gaps of both bulks ($E < \min(m_1,m_2)$), propagate along the $\hat{x}$-direction ($k_x \in \mathbb{R}$), and decay along the $y$-direction ($\kappa_1,\kappa_2 \in \mathbb{R}_+$).
The only physical solution is $\kappa_1 = m_1/v$, $\kappa_2 = m_2/v$, and $E = k_x v$.
For this solution, $t=-(m_2 + k_xv) / (m_1 - k_xv)$.

The SPC$^-$-VPC$^+$ ($\Delta_s < 0$ and $\Delta_v > 0$) interface is solved in the same way.
The valley $K$ and spin up polarized solution has the same decay constants and dispersion relation as Eq.~\ref{eq:step_2}.

Overall, the decay constant $\kappa$ is proportional to the half width of the band gap $m$.
Furthermore, the spin and valley DoFs of the edge mode remain unchanged when the band gap size changes.

\begin{figure}
    \centering
    \includegraphics[width=0.6\textwidth]{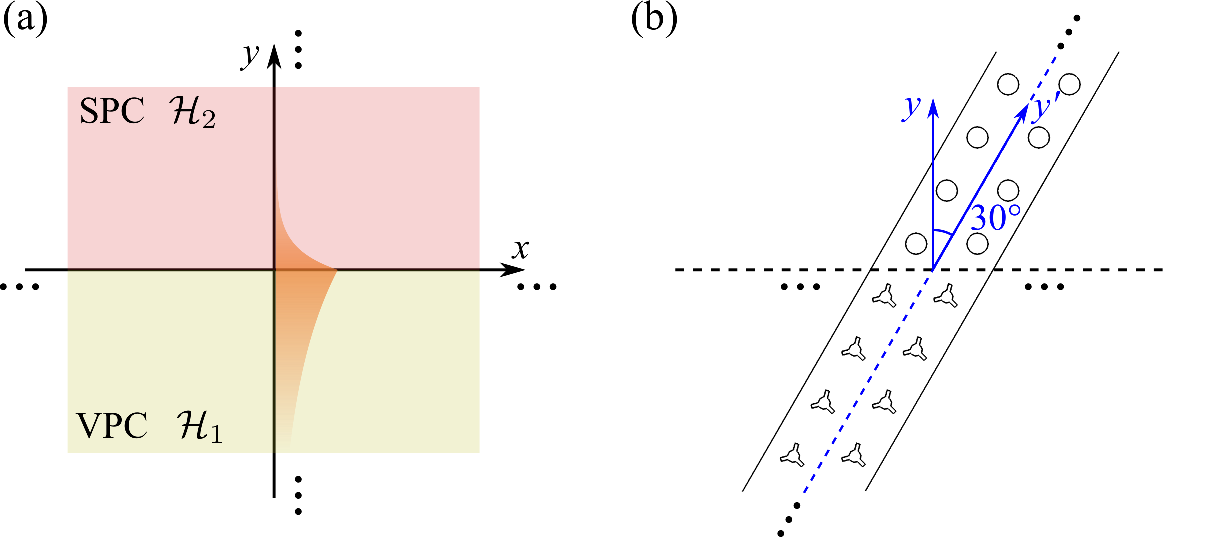}
    \caption{
    (a) The SPC-VPC interface.
    The VPC (SPC) spans the entire $y<0$ ($y>0$) half-plane.
    The edge mode (represented in orange) has different decay constants into the two domains.
    (b) The SPC-VPC interface used in COMSOL simulation.
    the energy density data is taken along the blue dashed line.
    The angle between the $\hat{y}'$-direction and the $\hat{y}$-direction is $30^{\circ}$.}
    %In this setup, the tripod angle is $\theta = 45^{\circ}$.
    \label{fig:interface}
\end{figure}

On the other hand, we evaluate the decay constant by fitting a decay function to COMSOL numerical data.
In a periodic lattice structure, the wave decaying along the transverse direction can be described as a Bloch state form for $y'>0$ (Fig.~\ref{fig:interface}~(b)),
\begin{equation}
\label{eq:transverse_decay}
\psi(y') \sim e^{-\kappa y'} u(y'),
\end{equation}
where $u(y')$ is a periodic function with the lattice constant $a_0$ as its periodicity, and $u(y')$ can be written as $\sum_n p_n \exp{(i 2\pi n y'/a_0)}$, $n \in \mathbb{Z}_+$, and $p_n$ are the coefficients.

We extract the COMSOL-calculated time-averaged energy density ($\langle U \rangle$) data at the SPC-VPC interface (Fig.~\ref{fig:interface}(b)) along a zigzag direction (the $\hat{y}'$-direction) that is $30^{\circ}$ from the transverse direction ($\hat{y}$-direction).
This choice is to avoid the proximity of metal, where the electromagnetic field changes drastically, affecting the quality of the fitting. To simplify the fitting procedure, we only consider the leading two terms ($n=0,1$) in Eq.~\ref{eq:transverse_decay} and fit the following function into the normalized time-averaged energy density for $y'>0$ (SPC domain) and $y'<0$ (VPC domain) separately,
\begin{equation}
\label{eq:fit_function}
e^{-2 \kappa |y'| \cos{(\pi / 6)}} [p_0 + p_1 \cos{(2\pi |y' - y_0|)}],
\end{equation}
where $\kappa$, $p_0$, $p_1$, and $y_0$ are fitting coefficients.
The factor of $2$ before $\kappa$ is because energy is proportional to the modulus square of the field ($\langle U \rangle \propto |\psi|^2$).
The $\cos{(\pi / 6)}$ term comes from projecting $y'$ to $y$ (Fig.~\ref{fig:interface}~(b)).

\begin{figure}
    \centering
    \includegraphics[width=0.5\textwidth]{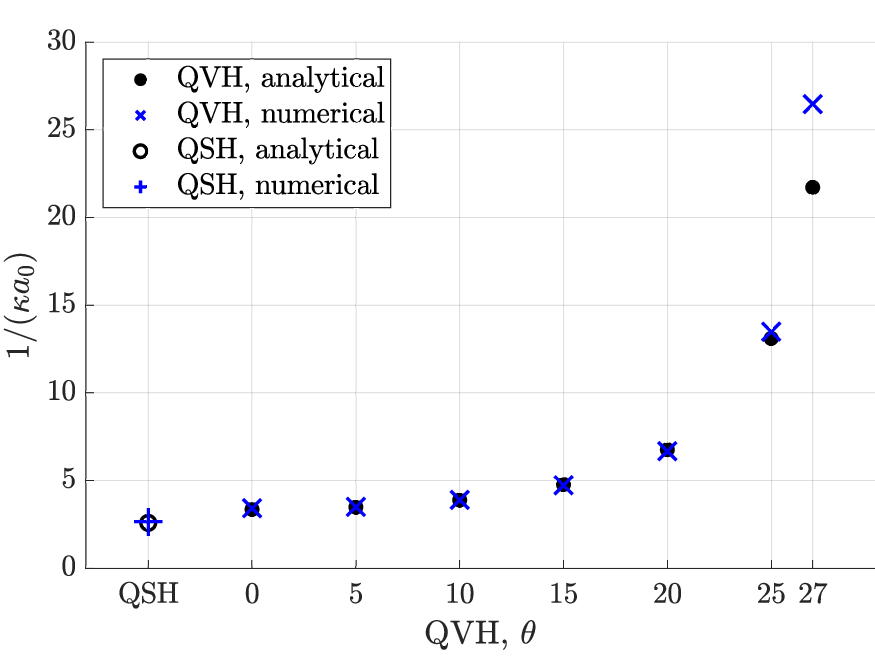}
    \caption{
    The inverse of the decay constant, measured in the number of unit cells.
    As $\theta$ changes from $0^{\circ}$ to $27^{\circ}$, the band gap narrows, and the exponential decay becomes more smooth.
    }
    \label{fig:compare_decay_const}
\end{figure}

\begin{figure}
  \centering
  \includegraphics[width=\textwidth]{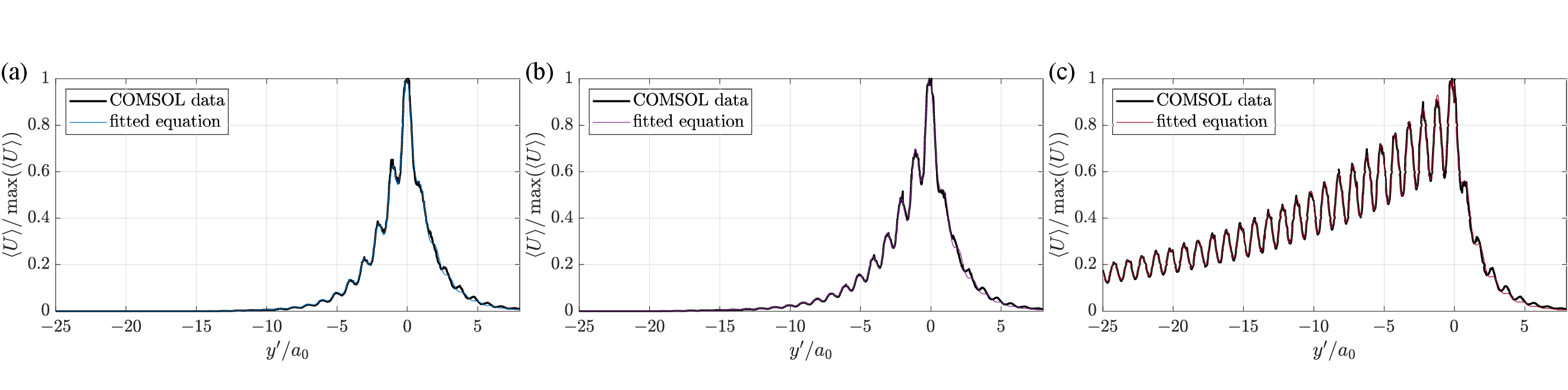}
  \caption{Time-averaged energy density along the $\hat{y}'$-direction in Fig.~\ref{fig:interface}(b) for $3$ different tripod angle configurations:
  (a) $\theta = 0^{\circ}$; (b) $\theta = 15^{\circ}$; (c) $\theta = 27^{\circ}$.
  With only two leading terms in $u(y)$, Eq.~\ref{eq:fit_function} provides sufficiently satisfactory fitting.
  For all $7$ interfaces, the numerically calculated $\kappa_2$ in the SPC domain are almost the same (deviation$<3.0\%$).}
  \label{fig:Wav_distribution_along_y'}
\end{figure}

The analytical and numerical results of the decay constants for the SPC and the VPC with $7$ different tripod configurations are presented in Fig.~\ref{fig:compare_decay_const}.
The quantity $1/(\kappa a_0)$ represents the localization distance where the amplitude of the field reduces to $1/e$ of its maximum value.
This distance is measured in the number of unit cells.
For $30^{\circ} < \theta < 60^{\circ}$, $\kappa$ can be calculated through $\kappa(\theta) = \kappa(60^{\circ} - \theta)$.
Fig.~\ref{fig:Wav_distribution_along_y'} shows three examples of the $\langle U \rangle$ distributions and the related fitting.

\section*{Finiteness along the transverse direction}
\begin{figure}
    \centering
    \includegraphics[width=0.7\textwidth]{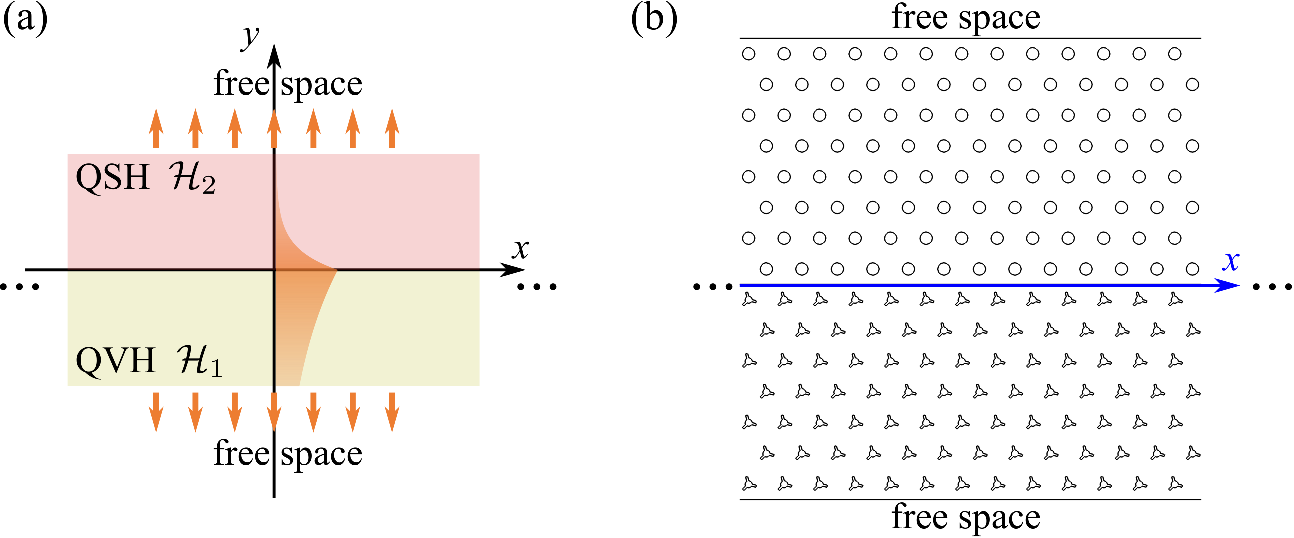}
    \caption{
    (a) The SPC-VPC interface.
    Both PTIs are finite along the $\hat{y}$-direction.
    The edge mode (represented in orange) out-couples to free space at the air-PTI interface, resulting in energy loss.
    (b) The finite SPC-VPC interface used in COMSOL simulation.
    The blue line represents the $\hat{x}$-direction, where the energy density data is taken.
    In this setup, the tripod angle $\theta = 45^{\circ}$.}
    \label{fig:decay_along_x}
\end{figure}

When the photonic crystal is finite along the transverse direction (Fig.~\ref{fig:decay_along_x}~(a)), the exponential decay tail at the air-PTI interface causes optical energy to out-couple into free space.
Therefore, the edge mode is also decaying along the longitudinal direction, because of the energy loss at the transverse directional boundaries.
The edge mode can be written as,
\begin{equation}
\label{eq:longitudinal_decay}
\psi(x,y) \propto
\begin{cases}
e^{ikx -\kappa_x x}e^{-\kappa_2 y}, & \text{if}\ y>0, \\
e^{ikx -\kappa_x x}e^{\kappa_1 y}, & \text{if}\ y<0.
\end{cases}
\end{equation}

\begin{figure}
  \centering
  \includegraphics[width=0.8\textwidth]{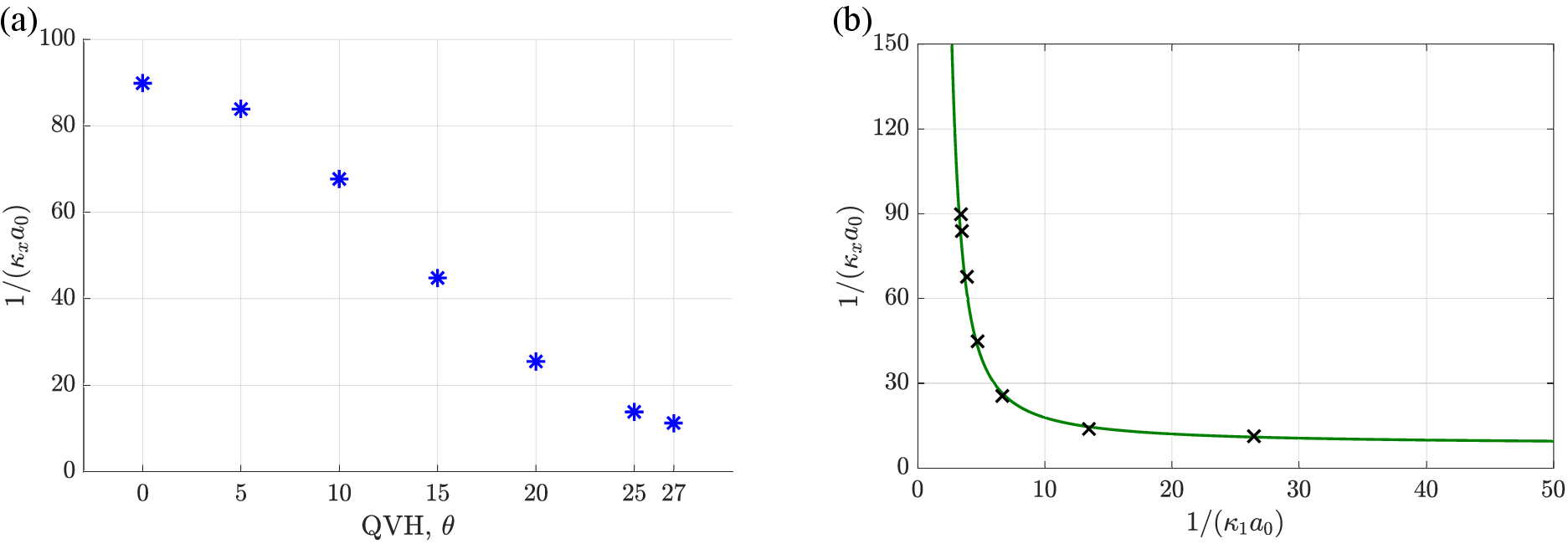}
  \caption{
  (a) The inverse of the decay constant along the $\hat{x}$-direction, measured in the number of unit cells.
  (b) The inverses of $\kappa_1$ and $\kappa_x$ and the fitted function $\kappa_x = \kappa_{x0} e^{-t \kappa_1}$, where $\kappa_{x0} \approx 0.12/a_0$ and $t \approx 7.9 a_0$.
  }
  \label{fig:decay_const_along_x}
\end{figure}

Qualitatively, $\kappa_x$ should be negatively correlated with $\kappa_{1,2}$, because the more rapid the mode decays along the $\hat{y}$-direction, the less energy tunnels into the free space, causing more energy to propagate along the $\hat{x}$-direction.
Quantitatively, however, because the out-coupling efficiency at the air-PTI interface depends on the geometrical detail of the PTI and the nature of the electromagnetic wave, without knowing the exact boundary condition, one cannot calculate $\kappa_x$.
Therefore, we find $\kappa_x$ through fitting an exponential decay curve to the COMSOL-calculated time-averaged energy density data of the edge mode along the longitudinal direction (Fig.~\ref{fig:decay_along_x}~(b)).
For all $7$ different tripod angle configurations, the inverse of $\kappa_x$ is measured in the number of unit cells and plotted in Fig.~\ref{fig:decay_const_along_x}.

Then, we investigate the relation between $\kappa_1$ and $\kappa_x$.
This relation cannot be derived with merely analytical approaches without knowing the exact boundary conditions.
Therefore, we first consider two extreme cases ($\kappa_1 \to \infty\ \text{or}\ 0$) and then choose a proper function to fit in the $(\kappa_1,\kappa_x)$ data for the $7$ different tripod orientations.

For the first case, $\kappa_1 \to \infty$, the exponential decay tail into the VPC domain is infinitely short, so, at the air-VPC interface, the amount of energy tunneled into the free space is zero.
Meanwhile, at the air-SPC interface ($y=+4\sqrt{3}a_0$), the field amplitude decays to $e^{- 4\sqrt{3} a_0 \kappa_2}$ of its original value.
Because $e^{- 4\sqrt{3} a_0 \kappa_2} \ll 1$ and the edge mode does not have a real-valued momentum along the $+\hat{y}$-direction, negligible amount of energy tunnels at the air-QSH interface.
Overall, when $\kappa_1 \to \infty$, the energy loss into the free space is negligible, so the edge mode does not attenuate along the longitudinal direction, i.e., $\kappa_x \to 0$.

For the second case, $\kappa_1 \to 0$, the edge mode is a plane wave in the QVH domain ($y<0$ half-plane).
Though the wave amplitude does not decay for $y<0$, the edge mode does not have a real-valued momentum along the $-\hat{y}$-direction, so the out-coupling at the air-QVH interface does not consume all the energy of the edge mode, i.e., is finite.
Meanwhile, at the air-QSH interface, the energy loss is still negligible.
Overall, when $\kappa_1 \to 0$, the energy loss efficiency along the transverse direction is not $100\%$, so the edge mode still has a finite attenuation along the longitudinal direction, i.e., $\kappa_x \not\to \infty$.

Concluding the two extreme cases, we find that the relation between $\kappa_1$ and $\kappa_x$ should satisfy,
\begin{equation}
\label{eq:decayxy_relation}
\lim_{\kappa_1 \to \infty} \kappa_x \to 0\ \text{and}\ 
\lim_{\kappa_1 \to 0} \kappa_x \to \kappa_{x0},\ 
\kappa_{x0} \in \mathbb{R}_+.
\end{equation}

A function that satisfies Eq.~\ref{eq:decayxy_relation} is $\kappa_x = \kappa_{x0} e^{-t \kappa_1}$.
The fitted function is plotted in Fig.~\ref{fig:decay_const_along_x}~(b).

\section*{Semi-analytical estimation of the ratio between the energy received at the two ports}
\begin{figure}
    \centering
    \includegraphics[width=0.8\textwidth]{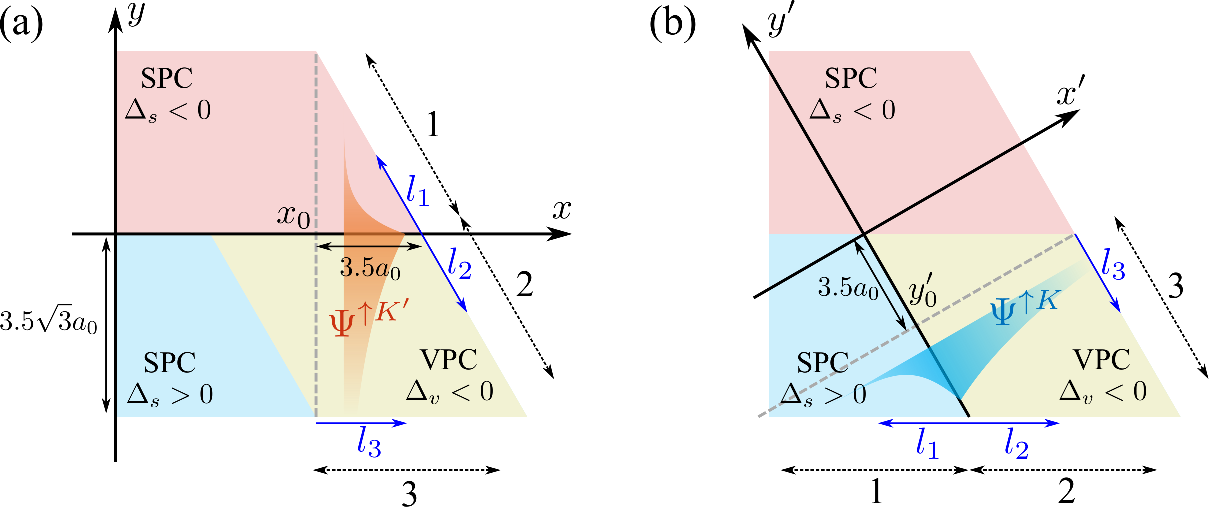}
    \caption{
    Schematic diagrams of the TREK states propagating in the PTI structure.
    (a) The $\Psi^{\uparrow K'}$ state.
    (b) The $\Psi^{\uparrow K}$ state.
    For the $\Psi^{\uparrow K}$ state, the coordinates in Eq.~\ref{eq:wave_at_x0} are changed: $x \rightarrow -y'$, $y \rightarrow -x'$.
    }
    \label{fig:integral}
\end{figure}

To estimate the ratio between the received energy at the two ports with semi-analytical techniques, we consider the interface between two continuums (Fig.~\ref{fig:integral}), which mimics our PTI structure.
Assuming the wave at point $x_0$ is
\begin{equation}
\label{eq:wave_at_x0}
\psi(x_0,y)=
\begin{cases}
A e^{(ik - \kappa_x) x_0} e^{-\kappa_2 y}, & \text{for}\ y>0, \\
A e^{(ik - \kappa_x) x_0} e^{\kappa_1 y}, & \text{for}\ y<0,
\end{cases}
\end{equation}
we can write the wave amplitude along the interface section $1$, $2$, and $3$ in Fig.~\ref{fig:integral}(a) as,
\begin{equation}
\begin{aligned}\label{eq:wave_at_sections}
& \psi_{\text{sec 1}} (l_1)
= A e^{(ik-\kappa_x) (3.5 a_0 + x_0 - l_1/2)} e^{-\sqrt{3}/2 \kappa_2 l_1}, \\
& \psi_{\text{sec 2}} (l_2)
= A e^{(ik-\kappa_x) (3.5 a_0 + x_0 + l_2/2)} e^{-\sqrt{3}/2 \kappa_1 l_2}, \\
& \psi_{\text{sec 3}} (l_3)
= A e^{(ik-\kappa_x) (x_0 + l_3)} e^{-7\sqrt{3}/2 \kappa_1 a_0},
\end{aligned}
\end{equation}
where section $1$ and $2$ constitute ``transmission" and $3$ is ``tunneling".
We have omitted the time-harmonics term $e^{-i\omega t}$, because it contributes $1$ to the modulus square.
We evaluate $\int dl |\psi|^2$ at the interface sections $1$, $2$, and $3$.
\begin{equation}
\label{eq:integrals}
\int dl |\psi|^2 =
\begin{cases}
\int_0^{6a_0} dl_1 |\psi_{\text{sec 1}} (l_1)|^2
= |A|^2 e^{-\kappa_x (2x_0 + 7a_0)} \dfrac{1 - e^{6a_0(\kappa_x - \sqrt{3}\kappa_2)}}{- \kappa_x + \sqrt{3}\kappa_2}, & \text{for section}\ 1, \\
\int_0^{6a_0} dl_2 |\psi_{\text{sec 2}} (l_2)|^2
= |A|^2 e^{-\kappa_x (2x_0 + 7a_0)} \dfrac{1 - e^{-6a_0(\kappa_x + \sqrt{3}\kappa_1)}}{\kappa_x + \sqrt{3}\kappa_1}, & \text{for section}\ 2, \\
\int_0^{6a_0} dl_3 |\psi_{\text{sec 3}} (l_3)|^2
= |A|^2 e^{-2\kappa_x x_0 -7\sqrt{3} \kappa_1 a_0} \dfrac{1 - e^{-12 \kappa_x a_0}}{2\kappa_x}, & \text{for section}\ 3.
\end{cases}
\end{equation}

For the spin up and valley $K'$ polarization, the semi-analytical description of the energy received at ports $C$ and $B$ are,
\begin{equation}
\begin{aligned}\label{eq:ports}
& W_C = \eta^{\uparrow K'} W_{tr} + \eta^{\uparrow K} W_{tu},\ W_B = \eta^{\uparrow K} W_{tr} + \eta^{\uparrow K'} W_{tu}, & \text{for}\ 0^{\circ}<\theta<30^{\circ}, \\
& W_C = \eta^{\uparrow K} W_{tr} + \eta^{\uparrow K'} W_{tu},\ W_B = \eta^{\uparrow K'} W_{tr} + \eta^{\uparrow K} W_{tu}, & \text{for}\ 30^{\circ}<\theta<60^{\circ},
\end{aligned}
\end{equation}
where $W_{tr} \propto \int_{\text{sec 1}} |\psi|^2 + \int_{\text{sec 2}} |\psi|^2$ and $W_{tu} \propto \int_{\text{sec 3}} |\psi|^2$ are the transmitted and tunneled energy.
$\eta^{\uparrow K'}$ ($\eta^{\uparrow K}$) is the excitation efficiency for the $\Psi^{\uparrow K'}$ ($\Psi^{\uparrow K}$) TREK state:
\begin{equation}
\label{eq:efficiency}
\eta^{s v} \propto |\mathbf{d} \cdot \mathbf{E}^{s v}(\omega, \mathbf{r})|^2,
\end{equation}
where $s=\uparrow,\downarrow$ and $v=K,K'$.
$\mathbf{d}$ is the electric dipole moment, and $\mathbf{E}^{s v}(\omega, \mathbf{r})$ is the electric field profile of the eigenmode $\Psi^{s v}$ at frequency $\omega$ and position $\mathbf{r}$~\cite{Lodahl:2015,Lodahl_Review:2015}.
In our setup, the dipole is $\hat{z}$-directional.
Therefore, the excitation efficiency can be simplified as,
\begin{equation}
\label{eq:efficiency 2}
\eta^{s v} \propto |E_z^{s v}(\omega, \mathbf{r})|^2.
\end{equation}

\section*{Excitation efficiency of the antenna source}
\begin{figure}
    \centering
    \includegraphics[width=0.8\textwidth]{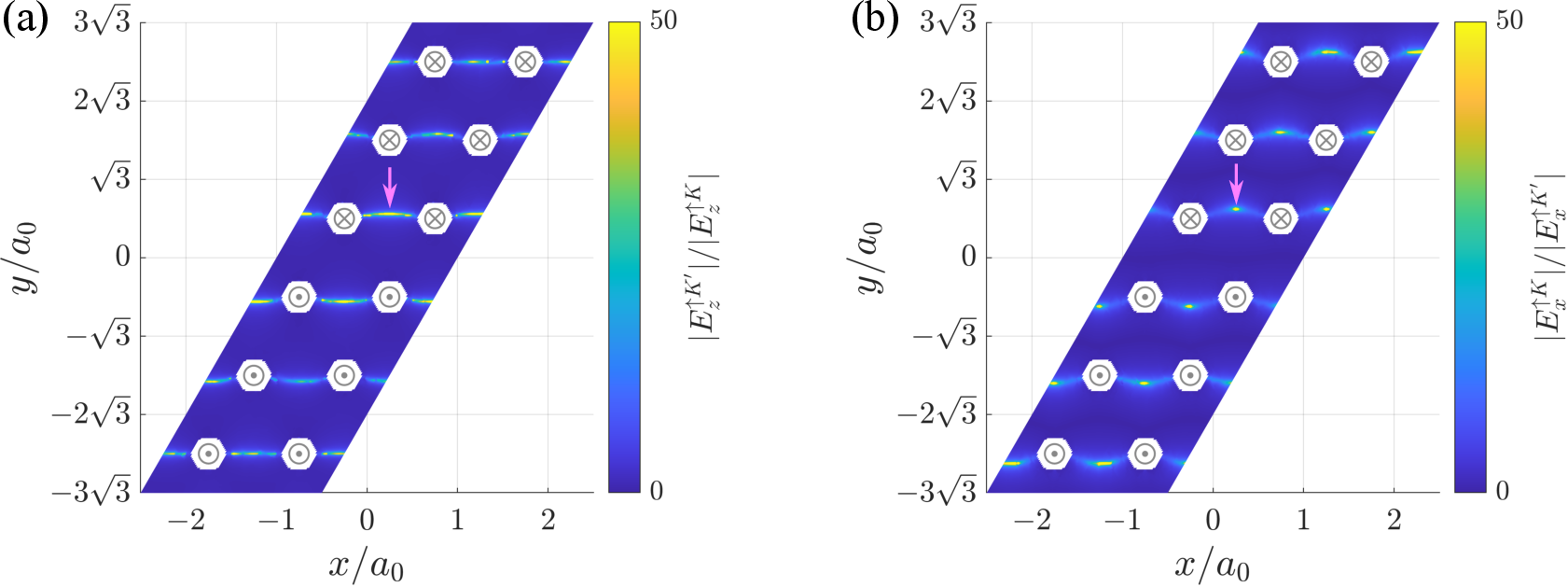}
    \caption{The ratio between the electric fields of the $K$ and $K'$ valley-polarizations for (a) $\hat{z}$-directional components, (b) $\hat{x}$-directional components.
    The plane is at $z=0$.}
    \label{fig:diverging_ratio}
\end{figure}

The excitation efficiency ratio $\eta^{\uparrow K'} / \eta^{\uparrow K}$ (or $\eta^{\uparrow K} / \eta^{\uparrow K'}$) can be considerably high if an ideally small source is used.
For example, at the spot marked by the magenta arrow in Fig.~\ref{fig:diverging_ratio}(a), $\eta^{\uparrow K'} / \eta^{\uparrow K} \equiv |E_z^{\uparrow K'}|^2 / |E_z^{\uparrow K}|^2 \approx 1\times 10^4$, indicating that the $K'$ valley-polarized state can be excited by a $\hat{z}$-directional electric dipole with significantly high preference.
With a $\hat{x}$-directional electric dipole placed at the marked position in Fig.~\ref{fig:diverging_ratio}(b), $\eta^{\uparrow K} / \eta^{\uparrow K'} \equiv |E_x^{\uparrow K}|^2 / |E_x^{\uparrow K'}|^2 \approx 4\times 10^4$, and the $K$ valley-polarized state can be excited with significantly high preference.

\begin{figure}
    \centering
    \includegraphics[width=1\textwidth]{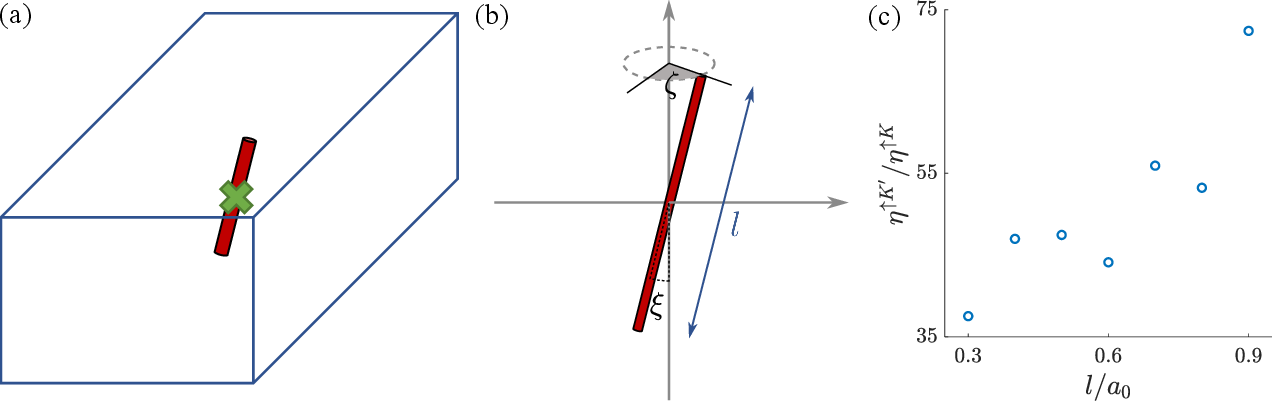}
    \caption{(a) Current segment (red line) in the SPC$^+$-SPC$^-$ waveguide, functioning as an excitation source.
    (b) The current segment of length $l$ is slightly tilted and rotated.
    The azimuthal angle $\zeta$ and polar angle $\xi$ that satisfy $0<\zeta<2\pi$ and $0<\xi<\pi/18$ are uniformly distributed random variables.
    The mid-point of the current segment is deviated by $\delta r$ along $x$, $y$, and $z$ directions from the position labeled by the green cross [corresponding to the position in main text Fig.~2(a)], where $\delta r$ is a random number satisfying $-0.05a_0<\delta r<0.05a_0$. ($0.05a_0 \approx 1.8$mm in the microwave experiment.)
    (c) The calculated excitation ratio of such random antennae with different lengths $l$.
    Each point is the result averaged over $100$ random antennae of the same length.}
    \label{fig:random_antenna}
\end{figure}

To carefully estimate the excitation efficiency of the current probe used in our microwave experiments, we calculate the excitation efficiency of a current source with distribution $\mathbf{J}(\mathbf{r})$.
The excited field amplitude $A$ satisfies $A \propto \iiint_V \mathbf{E}(\mathbf{r}) \cdot \mathbf{J}(\mathbf{r}) dV$, where $\mathbf{E}(\mathbf{r})$ is the electric field distribution of the excited eigen mode~\cite{Pozar}.
The excitation efficiency satisfies $\eta \propto |A|^2$.
A $\hat{z}$-directional current segment can be represented as
\begin{equation}
\mathbf{J}(\mathbf{r}) = j_0 \delta(x-x_0) \delta(y-y_0) [\Theta(z-l_0/2)-\Theta(z+l_0/2)] \hat{z},
\end{equation}
where $\Theta(z)$ is the Heaviside step function.

To take into consideration possible slight misalignments happened in experiment, we let the current segment be slightly and randomly tilted and rotated.
Also, its mid-point is deviated from the designed position with a small, random deviation (see Fig.~\ref{fig:random_antenna} for details).
We vary the antenna length and perform this numerical procedure to study how the excitation ratio depends on the antenna length.
The result is shown in Fig.~\ref{fig:random_antenna}(c).
Generally, a longer antenna provides a slightly higher excitation ratio.

\section*{The experimental setup}
\begin{figure}
    \centering
    \includegraphics[width=1\textwidth]{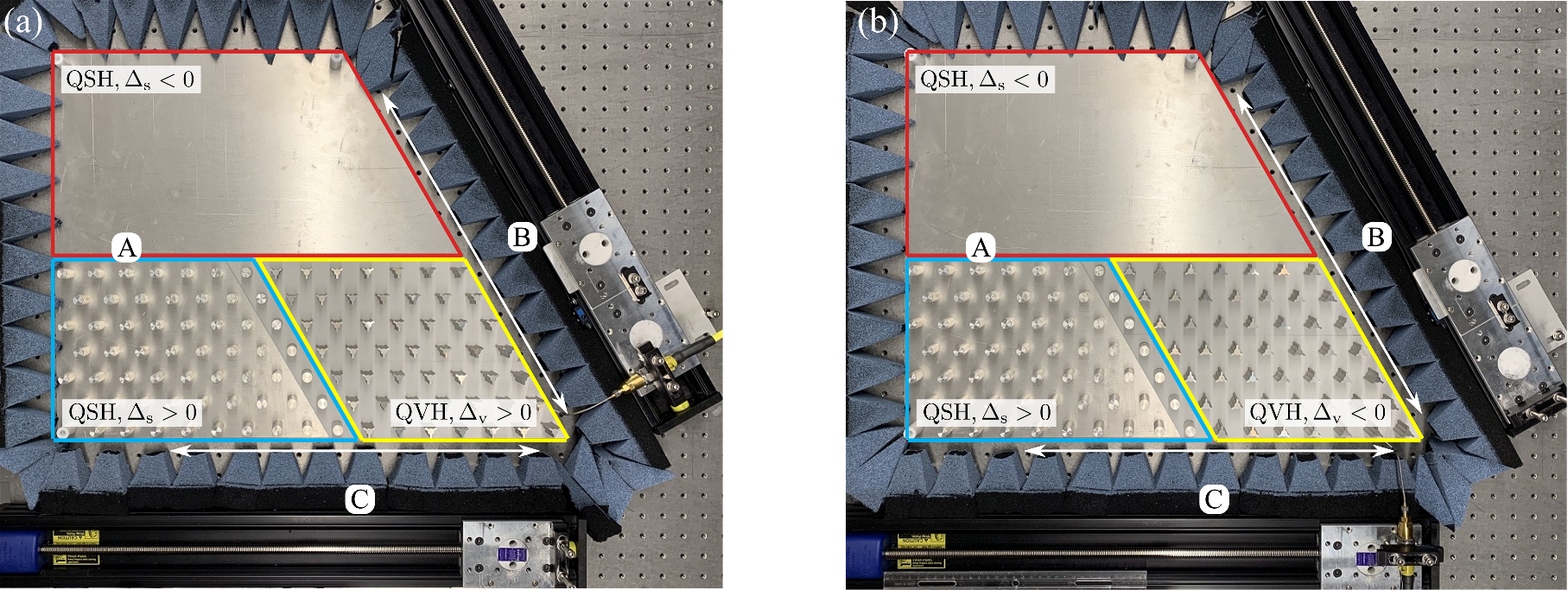}
    \caption{The manufactured structure used in experiment.
    (a) tripod angle $\theta = 0^{\circ}$, corresponding to $\Delta_v > 0$;
    (b) tripod angle $\theta = 60^{\circ}$, corresponding to $\Delta_v < 0$.
    The aluminum rods of the $\Delta_s < 0$ SPC are not shown, because they are assembled on the top plate which is lifted for visualization.
    The structure is surrounded by electromagnetic wave absorbers to eliminate the undesired reflection caused by the external metallic structure.}
    \label{fig:structure}
\end{figure}
Fig.~\ref{fig:structure} shows the structure used in the microwave experiment.
The aluminum tripods used to build the VPC are fabricated using wire electrical discharge machining.
Two pieces of structural foam (ROHACELL 51HF) are used to support the tripods between the two parallel plates.
Another piece of structural foam with laser-cut holes is used for precisely locating the position and the directional angle of every tripod.
The source is a $23$mm long, $\hat{z}$-directional antenna.

We use a vector network analyzer (Keysight N5222A) to perform the measurement.
Each set of measurement contains $257$ transmission spectra taken at uniformly spaced positions $r_i$ along periphery B or C of structure. To calculate the total amount of detected energy at a port, we sum $|S_{21}(r_i)|^2$ over all $r_i$ along that periphery.
The detect antenna is mounted on a motor-driven Velmex BiSlide rail.
Electromagnetic wave absorbers are attached to one side of the rail to prevent reflection from the metallic surface.

\section*{Embedding information in valley-polarized light}
\begin{figure}
    \centering
    \includegraphics[width=0.9\textwidth]{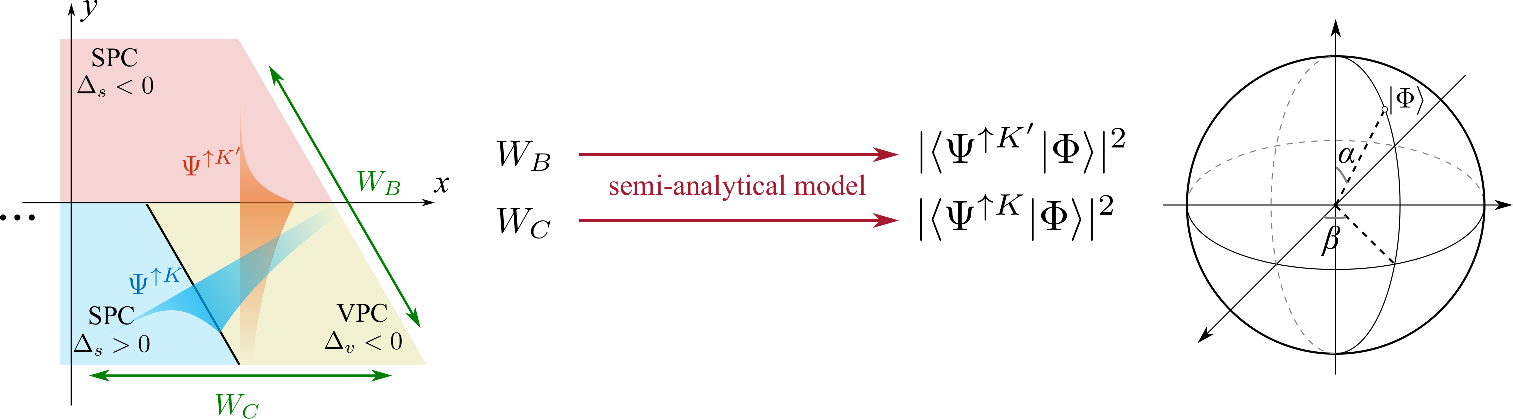}
    \caption{The semi-analytical model links the projections  $|\langle \Psi^{\uparrow K'} | \Phi \rangle|^2$ and $|\langle \Psi^{\uparrow K} | \Phi \rangle|^2$ with the measurement results $W_B$ and $W_C$.}
    \label{fig:valley_spinor}
\end{figure}

Any arbitrary spin-up state in the SPC$^+$-SPC$^-$ waveguide can be written as $\Phi = \cos(\alpha/2) \Psi^{\uparrow K} + e^{i \beta} \sin(\alpha/2) \Psi^{\uparrow K'}$.
This state can represent the bit $(\cos \alpha/2, e^{i \beta} \sin \alpha/2)^T$.
To decode the information represented by $\Phi$, one needs to measure the following two projections,
\begin{equation}
|\langle \Psi^{\uparrow K} | \Phi \rangle|^2 = \cos^2 (\alpha/2)
\text{ and }
|\langle \Psi^{\uparrow K'} | \Phi \rangle|^2 = \sin^2 (\alpha/2).
\end{equation}

However, because the topological band gap of the photonic crystal only spans a finite frequency range, the waveguide modes tunnel along the transverse direction.
Additionally, in most practical scenarios, the size of the structure is limited.
Consequently, the tunneling can compromise the measurement of the two projections, for example, when $30^{\circ} < \theta < 60^{\circ}$, $W_B / (W_B+W_C) \neq |\langle \Psi^{\uparrow K'} | \Phi \rangle|^2$ and $W_C / (W_B+W_C) \neq |\langle \Psi^{\uparrow K} | \Phi \rangle|^2$.

With the semi-analytical model (Eq.~\ref{eq:ports}), one can calculate $|\langle \Psi^{\uparrow K'} | \Phi \rangle|^2$ and $|\langle \Psi^{\uparrow K} | \Phi \rangle|^2$ from the experimentally measured quantities $W_B$ and $W_C$, the localization distance $L_{\rm loc}$, and the structural dimension (Fig.~\ref{fig:valley_spinor}).
Hence, this problem can be resolved without an ideally large photonic crystal.

\section*{References}
\bibliographystyle{unsrt}

\section*{References}
\begin{thebibliography}{41}%
\makeatletter
\providecommand \@ifxundefined [1]{%
 \@ifx{#1\undefined}
}%
\providecommand \@ifnum [1]{%
 \ifnum #1\expandafter \@firstoftwo
 \else \expandafter \@secondoftwo
 \fi
}%
\providecommand \@ifx [1]{%
 \ifx #1\expandafter \@firstoftwo
 \else \expandafter \@secondoftwo
 \fi
}%
\providecommand \natexlab [1]{#1}%
\providecommand \enquote  [1]{``#1''}%
\providecommand \bibnamefont  [1]{#1}%
\providecommand \bibfnamefont [1]{#1}%
\providecommand \citenamefont [1]{#1}%
\providecommand \href@noop [0]{\@secondoftwo}%
\providecommand \href [0]{\begingroup \@sanitize@url \@href}%
\providecommand \@href[1]{\@@startlink{#1}\@@href}%
\providecommand \@@href[1]{\endgroup#1\@@endlink}%
\providecommand \@sanitize@url [0]{\catcode `\\12\catcode `\$12\catcode
  `\&12\catcode `\#12\catcode `\^12\catcode `\_12\catcode `\%12\relax}%
\providecommand \@@startlink[1]{}%
\providecommand \@@endlink[0]{}%
\providecommand \url  [0]{\begingroup\@sanitize@url \@url }%
\providecommand \@url [1]{\endgroup\@href {#1}{\urlprefix }}%
\providecommand \urlprefix  [0]{URL }%
\providecommand \Eprint [0]{\href }%
\providecommand \doibase [0]{http://dx.doi.org/}%
\providecommand \selectlanguage [0]{\@gobble}%
\providecommand \bibinfo  [0]{\@secondoftwo}%
\providecommand \bibfield  [0]{\@secondoftwo}%
\providecommand \translation [1]{[#1]}%
\providecommand \BibitemOpen [0]{}%
\providecommand \bibitemStop [0]{}%
\providecommand \bibitemNoStop [0]{.\EOS\space}%
\providecommand \EOS [0]{\spacefactor3000\relax}%
\providecommand \BibitemShut  [1]{\csname bibitem#1\endcsname}%
\let\auto@bib@innerbib\@empty
%</preamble>
\bibitem [{\citenamefont {Bogaerts}\ \emph {et~al.}(2020)\citenamefont
  {Bogaerts}, \citenamefont {P{\'{e}}rez}, \citenamefont {Capmany},
  \citenamefont {Miller}, \citenamefont {Poon}, \citenamefont {Englund},
  \citenamefont {Morichetti},\ and\ \citenamefont {Melloni}}]{PIC_Review:2020}%
  \BibitemOpen
  \bibfield  {author} {\bibinfo {author} {\bibfnamefont {W.}~\bibnamefont
  {Bogaerts}}, \bibinfo {author} {\bibfnamefont {D.}~\bibnamefont
  {P{\'{e}}rez}}, \bibinfo {author} {\bibfnamefont {J.}~\bibnamefont
  {Capmany}}, \bibinfo {author} {\bibfnamefont {D.~A.~B.}\ \bibnamefont
  {Miller}}, \bibinfo {author} {\bibfnamefont {J.}~\bibnamefont {Poon}},
  \bibinfo {author} {\bibfnamefont {D.}~\bibnamefont {Englund}}, \bibinfo
  {author} {\bibfnamefont {F.}~\bibnamefont {Morichetti}}, \ and\ \bibinfo
  {author} {\bibfnamefont {A.}~\bibnamefont {Melloni}},\ }\bibfield  {title}
  {\enquote {\bibinfo {title} {Programmable photonic circuits},}\ }\href
  {\doibase 10.1038/s41586-020-2764-0} {\bibfield  {journal} {\bibinfo
  {journal} {Nature}\ }\textbf {\bibinfo {volume} {586}},\ \bibinfo {pages}
  {207--216} (\bibinfo {year} {2020})}\BibitemShut {NoStop}%
\bibitem [{\citenamefont {Capmany}\ and\ \citenamefont
  {P{\'{e}}rez}(2020)}]{PIC_Book}%
  \BibitemOpen
  \bibfield  {author} {\bibinfo {author} {\bibfnamefont {J.}~\bibnamefont
  {Capmany}}\ and\ \bibinfo {author} {\bibfnamefont {D.}~\bibnamefont
  {P{\'{e}}rez}},\ }\href {\doibase 10.1093/oso/9780198844402.001.0001} {\emph
  {\bibinfo {title} {Programmable Integrated Photonics}}}\ (\bibinfo
  {publisher} {Oxford University Press},\ \bibinfo {year} {2020})\BibitemShut
  {NoStop}%
\bibitem [{\citenamefont {Marpaung}, \citenamefont {Yao},\ and\ \citenamefont
  {Capmany}(2019)}]{PIC_Review:2019}%
  \BibitemOpen
  \bibfield  {author} {\bibinfo {author} {\bibfnamefont {D.}~\bibnamefont
  {Marpaung}}, \bibinfo {author} {\bibfnamefont {J.}~\bibnamefont {Yao}}, \
  and\ \bibinfo {author} {\bibfnamefont {J.}~\bibnamefont {Capmany}},\
  }\bibfield  {title} {\enquote {\bibinfo {title} {Integrated microwave
  photonics},}\ }\href {\doibase 10.1038/s41566-018-0310-5} {\bibfield
  {journal} {\bibinfo  {journal} {Nature Photonics}\ }\textbf {\bibinfo
  {volume} {13}},\ \bibinfo {pages} {80--90} (\bibinfo {year}
  {2019})}\BibitemShut {NoStop}%
\bibitem [{\citenamefont {Agrell}\ \emph {et~al.}(2016)\citenamefont {Agrell},
  \citenamefont {Karlsson}, \citenamefont {Chraplyvy}, \citenamefont
  {Richardson}, \citenamefont {Krummrich}, \citenamefont {Winzer},
  \citenamefont {Roberts}, \citenamefont {Fischer}, \citenamefont {Savory},
  \citenamefont {Eggleton}, \citenamefont {Secondini}, \citenamefont
  {Kschischang}, \citenamefont {Lord}, \citenamefont {Prat}, \citenamefont
  {Tomkos}, \citenamefont {Bowers}, \citenamefont {Srinivasan}, \citenamefont
  {Brandt-Pearce},\ and\ \citenamefont {Gisin}}]{communication_Review:2016}%
  \BibitemOpen
  \bibfield  {author} {\bibinfo {author} {\bibfnamefont {E.}~\bibnamefont
  {Agrell}}, \bibinfo {author} {\bibfnamefont {M.}~\bibnamefont {Karlsson}},
  \bibinfo {author} {\bibfnamefont {A.~R.}\ \bibnamefont {Chraplyvy}}, \bibinfo
  {author} {\bibfnamefont {D.~J.}\ \bibnamefont {Richardson}}, \bibinfo
  {author} {\bibfnamefont {P.~M.}\ \bibnamefont {Krummrich}}, \bibinfo {author}
  {\bibfnamefont {P.}~\bibnamefont {Winzer}}, \bibinfo {author} {\bibfnamefont
  {K.}~\bibnamefont {Roberts}}, \bibinfo {author} {\bibfnamefont {J.~K.}\
  \bibnamefont {Fischer}}, \bibinfo {author} {\bibfnamefont {S.~J.}\
  \bibnamefont {Savory}}, \bibinfo {author} {\bibfnamefont {B.~J.}\
  \bibnamefont {Eggleton}}, \bibinfo {author} {\bibfnamefont {M.}~\bibnamefont
  {Secondini}}, \bibinfo {author} {\bibfnamefont {F.~R.}\ \bibnamefont
  {Kschischang}}, \bibinfo {author} {\bibfnamefont {A.}~\bibnamefont {Lord}},
  \bibinfo {author} {\bibfnamefont {J.}~\bibnamefont {Prat}}, \bibinfo {author}
  {\bibfnamefont {I.}~\bibnamefont {Tomkos}}, \bibinfo {author} {\bibfnamefont
  {J.~E.}\ \bibnamefont {Bowers}}, \bibinfo {author} {\bibfnamefont
  {S.}~\bibnamefont {Srinivasan}}, \bibinfo {author} {\bibfnamefont
  {M.}~\bibnamefont {Brandt-Pearce}}, \ and\ \bibinfo {author} {\bibfnamefont
  {N.}~\bibnamefont {Gisin}},\ }\bibfield  {title} {\enquote {\bibinfo {title}
  {Roadmap of optical communications},}\ }\href {\doibase
  10.1088/2040-8978/18/6/063002} {\bibfield  {journal} {\bibinfo  {journal}
  {Journal of Optics}\ }\textbf {\bibinfo {volume} {18}},\ \bibinfo {pages}
  {063002} (\bibinfo {year} {2016})}\BibitemShut {NoStop}%
\bibitem [{\citenamefont {Yang}\ \emph {et~al.}(2021)\citenamefont {Yang},
  \citenamefont {White}, \citenamefont {Ashtiani}, \citenamefont {Song},
  \citenamefont {Chang}, \citenamefont {Zou}, \citenamefont {Zhou},
  \citenamefont {Pang}, \citenamefont {Netherton}, \citenamefont {Ahn},
  \citenamefont {Skarda}, \citenamefont {Guidry}, \citenamefont {Su},
  \citenamefont {Vercruysse}, \citenamefont {Maclean}, \citenamefont
  {Aghaeimeibodi}, \citenamefont {Miller}, \citenamefont {Bowers},
  \citenamefont {Willner}, \citenamefont {Aflatouni},\ and\ \citenamefont
  {Vu\v{c}kovi\'{c}}}]{Vuckovic:2021}%
  \BibitemOpen
  \bibfield  {author} {\bibinfo {author} {\bibfnamefont {K.~Y.}\ \bibnamefont
  {Yang}}, \bibinfo {author} {\bibfnamefont {A.~D.}\ \bibnamefont {White}},
  \bibinfo {author} {\bibfnamefont {F.}~\bibnamefont {Ashtiani}}, \bibinfo
  {author} {\bibfnamefont {H.}~\bibnamefont {Song}}, \bibinfo {author}
  {\bibfnamefont {L.}~\bibnamefont {Chang}}, \bibinfo {author} {\bibfnamefont
  {K.}~\bibnamefont {Zou}}, \bibinfo {author} {\bibfnamefont {H.}~\bibnamefont
  {Zhou}}, \bibinfo {author} {\bibfnamefont {K.}~\bibnamefont {Pang}}, \bibinfo
  {author} {\bibfnamefont {A.}~\bibnamefont {Netherton}}, \bibinfo {author}
  {\bibfnamefont {G.~H.}\ \bibnamefont {Ahn}}, \bibinfo {author} {\bibfnamefont
  {J.~L.}\ \bibnamefont {Skarda}}, \bibinfo {author} {\bibfnamefont {M.~A.}\
  \bibnamefont {Guidry}}, \bibinfo {author} {\bibfnamefont {L.}~\bibnamefont
  {Su}}, \bibinfo {author} {\bibfnamefont {D.}~\bibnamefont {Vercruysse}},
  \bibinfo {author} {\bibfnamefont {J.-P.~W.}\ \bibnamefont {Maclean}},
  \bibinfo {author} {\bibfnamefont {S.}~\bibnamefont {Aghaeimeibodi}}, \bibinfo
  {author} {\bibfnamefont {D.~A.~B.}\ \bibnamefont {Miller}}, \bibinfo {author}
  {\bibfnamefont {J.~E.}\ \bibnamefont {Bowers}}, \bibinfo {author}
  {\bibfnamefont {A.~E.}\ \bibnamefont {Willner}}, \bibinfo {author}
  {\bibfnamefont {F.}~\bibnamefont {Aflatouni}}, \ and\ \bibinfo {author}
  {\bibfnamefont {J.}~\bibnamefont {Vu\v{c}kovi\'{c}}},\ }\href@noop {}
  {\enquote {\bibinfo {title} {Inverse-designed multi-dimensional silicon
  photonic transmitters},}\ } (\bibinfo {year} {2021}),\ \Eprint
  {http://arxiv.org/abs/2103.14139} {arXiv:2103.14139 [physics.app-ph]}
  \BibitemShut {NoStop}%
\bibitem [{\citenamefont {Li}, \citenamefont {Liu},\ and\ \citenamefont
  {Dai}(2018)}]{multimode_Review:2018}%
  \BibitemOpen
  \bibfield  {author} {\bibinfo {author} {\bibfnamefont {C.}~\bibnamefont
  {Li}}, \bibinfo {author} {\bibfnamefont {D.}~\bibnamefont {Liu}}, \ and\
  \bibinfo {author} {\bibfnamefont {D.}~\bibnamefont {Dai}},\ }\bibfield
  {title} {\enquote {\bibinfo {title} {Multimode silicon photonics},}\ }\href
  {\doibase 10.1515/nanoph-2018-0161} {\bibfield  {journal} {\bibinfo
  {journal} {Nanophotonics}\ }\textbf {\bibinfo {volume} {8}},\ \bibinfo
  {pages} {227--247} (\bibinfo {year} {2018})}\BibitemShut {NoStop}%
\bibitem [{\citenamefont {Bozinovic}\ \emph {et~al.}(2013)\citenamefont
  {Bozinovic}, \citenamefont {Yue}, \citenamefont {Ren}, \citenamefont {Tur},
  \citenamefont {Kristensen}, \citenamefont {Huang}, \citenamefont {Willner},\
  and\ \citenamefont {Ramachandran}}]{Bozinovic:2013}%
  \BibitemOpen
  \bibfield  {author} {\bibinfo {author} {\bibfnamefont {N.}~\bibnamefont
  {Bozinovic}}, \bibinfo {author} {\bibfnamefont {Y.}~\bibnamefont {Yue}},
  \bibinfo {author} {\bibfnamefont {Y.}~\bibnamefont {Ren}}, \bibinfo {author}
  {\bibfnamefont {M.}~\bibnamefont {Tur}}, \bibinfo {author} {\bibfnamefont
  {P.}~\bibnamefont {Kristensen}}, \bibinfo {author} {\bibfnamefont
  {H.}~\bibnamefont {Huang}}, \bibinfo {author} {\bibfnamefont {A.~E.}\
  \bibnamefont {Willner}}, \ and\ \bibinfo {author} {\bibfnamefont
  {S.}~\bibnamefont {Ramachandran}},\ }\bibfield  {title} {\enquote {\bibinfo
  {title} {Terabit-scale orbital angular momentum mode division multiplexing in
  fibers},}\ }\href {\doibase 10.1126/science.1237861} {\bibfield  {journal}
  {\bibinfo  {journal} {Science}\ }\textbf {\bibinfo {volume} {340}},\ \bibinfo
  {pages} {1545--1548} (\bibinfo {year} {2013})}\BibitemShut {NoStop}%
\bibitem [{\citenamefont {Moody}\ \emph {et~al.}(2021)\citenamefont {Moody},
  \citenamefont {Sorger}, \citenamefont {Juodawlkis}, \citenamefont {Loh},
  \citenamefont {Sorace-Agaskar}, \citenamefont {Davanco}, \citenamefont
  {Chang}, \citenamefont {Bowers}, \citenamefont {Quack}, \citenamefont
  {Galland}, \citenamefont {Aharonovich}, \citenamefont {Wolff}, \citenamefont
  {Schuck}, \citenamefont {Sinclair}, \citenamefont {Lončar}, \citenamefont
  {Komljenovic}, \citenamefont {Weld}, \citenamefont {Mookherjea},
  \citenamefont {Buckley}, \citenamefont {Radulaski}, \citenamefont
  {Reitzenstein}, \citenamefont {Pingault}, \citenamefont {Machielse},
  \citenamefont {Mukhopadhyay}, \citenamefont {Akimov}, \citenamefont
  {Zheltikov}, \citenamefont {Agarwal}, \citenamefont {Srinivasan},
  \citenamefont {Lu}, \citenamefont {Tang}, \citenamefont {Jiang},
  \citenamefont {McKenna}, \citenamefont {Safavi-Naeini}, \citenamefont
  {Steinhauer}, \citenamefont {Elshaari}, \citenamefont {Zwiller},
  \citenamefont {Davids}, \citenamefont {Martinez}, \citenamefont {Gehl},
  \citenamefont {Chiaverini}, \citenamefont {Mehta}, \citenamefont {Romero},
  \citenamefont {Lingaraju}, \citenamefont {Weiner}, \citenamefont {Peace},
  \citenamefont {Cernansky}, \citenamefont {Lobino}, \citenamefont {Diamanti},
  \citenamefont {Vidarte},\ and\ \citenamefont {Camacho}}]{QPIC_Review:2021}%
  \BibitemOpen
  \bibfield  {author} {\bibinfo {author} {\bibfnamefont {G.}~\bibnamefont
  {Moody}}, \bibinfo {author} {\bibfnamefont {V.~J.}\ \bibnamefont {Sorger}},
  \bibinfo {author} {\bibfnamefont {P.~W.}\ \bibnamefont {Juodawlkis}},
  \bibinfo {author} {\bibfnamefont {W.}~\bibnamefont {Loh}}, \bibinfo {author}
  {\bibfnamefont {C.}~\bibnamefont {Sorace-Agaskar}}, \bibinfo {author}
  {\bibfnamefont {M.}~\bibnamefont {Davanco}}, \bibinfo {author} {\bibfnamefont
  {L.}~\bibnamefont {Chang}}, \bibinfo {author} {\bibfnamefont {J.~E.}\
  \bibnamefont {Bowers}}, \bibinfo {author} {\bibfnamefont {N.}~\bibnamefont
  {Quack}}, \bibinfo {author} {\bibfnamefont {C.}~\bibnamefont {Galland}},
  \bibinfo {author} {\bibfnamefont {I.}~\bibnamefont {Aharonovich}}, \bibinfo
  {author} {\bibfnamefont {M.~A.}\ \bibnamefont {Wolff}}, \bibinfo {author}
  {\bibfnamefont {C.}~\bibnamefont {Schuck}}, \bibinfo {author} {\bibfnamefont
  {N.}~\bibnamefont {Sinclair}}, \bibinfo {author} {\bibfnamefont
  {M.}~\bibnamefont {Lončar}}, \bibinfo {author} {\bibfnamefont
  {T.}~\bibnamefont {Komljenovic}}, \bibinfo {author} {\bibfnamefont
  {D.}~\bibnamefont {Weld}}, \bibinfo {author} {\bibfnamefont {S.}~\bibnamefont
  {Mookherjea}}, \bibinfo {author} {\bibfnamefont {S.}~\bibnamefont {Buckley}},
  \bibinfo {author} {\bibfnamefont {M.}~\bibnamefont {Radulaski}}, \bibinfo
  {author} {\bibfnamefont {S.}~\bibnamefont {Reitzenstein}}, \bibinfo {author}
  {\bibfnamefont {B.}~\bibnamefont {Pingault}}, \bibinfo {author}
  {\bibfnamefont {B.}~\bibnamefont {Machielse}}, \bibinfo {author}
  {\bibfnamefont {D.}~\bibnamefont {Mukhopadhyay}}, \bibinfo {author}
  {\bibfnamefont {A.}~\bibnamefont {Akimov}}, \bibinfo {author} {\bibfnamefont
  {A.}~\bibnamefont {Zheltikov}}, \bibinfo {author} {\bibfnamefont {G.~S.}\
  \bibnamefont {Agarwal}}, \bibinfo {author} {\bibfnamefont {K.}~\bibnamefont
  {Srinivasan}}, \bibinfo {author} {\bibfnamefont {J.}~\bibnamefont {Lu}},
  \bibinfo {author} {\bibfnamefont {H.~X.}\ \bibnamefont {Tang}}, \bibinfo
  {author} {\bibfnamefont {W.}~\bibnamefont {Jiang}}, \bibinfo {author}
  {\bibfnamefont {T.~P.}\ \bibnamefont {McKenna}}, \bibinfo {author}
  {\bibfnamefont {A.~H.}\ \bibnamefont {Safavi-Naeini}}, \bibinfo {author}
  {\bibfnamefont {S.}~\bibnamefont {Steinhauer}}, \bibinfo {author}
  {\bibfnamefont {A.~W.}\ \bibnamefont {Elshaari}}, \bibinfo {author}
  {\bibfnamefont {V.}~\bibnamefont {Zwiller}}, \bibinfo {author} {\bibfnamefont
  {P.~S.}\ \bibnamefont {Davids}}, \bibinfo {author} {\bibfnamefont
  {N.}~\bibnamefont {Martinez}}, \bibinfo {author} {\bibfnamefont
  {M.}~\bibnamefont {Gehl}}, \bibinfo {author} {\bibfnamefont {J.}~\bibnamefont
  {Chiaverini}}, \bibinfo {author} {\bibfnamefont {K.~K.}\ \bibnamefont
  {Mehta}}, \bibinfo {author} {\bibfnamefont {J.}~\bibnamefont {Romero}},
  \bibinfo {author} {\bibfnamefont {N.~B.}\ \bibnamefont {Lingaraju}}, \bibinfo
  {author} {\bibfnamefont {A.~M.}\ \bibnamefont {Weiner}}, \bibinfo {author}
  {\bibfnamefont {D.}~\bibnamefont {Peace}}, \bibinfo {author} {\bibfnamefont
  {R.}~\bibnamefont {Cernansky}}, \bibinfo {author} {\bibfnamefont
  {M.}~\bibnamefont {Lobino}}, \bibinfo {author} {\bibfnamefont
  {E.}~\bibnamefont {Diamanti}}, \bibinfo {author} {\bibfnamefont {L.~T.}\
  \bibnamefont {Vidarte}}, \ and\ \bibinfo {author} {\bibfnamefont {R.~M.}\
  \bibnamefont {Camacho}},\ }\href@noop {} {\enquote {\bibinfo {title} {Roadmap
  on integrated quantum photonics},}\ } (\bibinfo {year} {2021}),\ \Eprint
  {http://arxiv.org/abs/2102.03323} {arXiv:2102.03323 [quant-ph]} \BibitemShut
  {NoStop}%
\bibitem [{\citenamefont {Bouchard}\ \emph {et~al.}(2021)\citenamefont
  {Bouchard}, \citenamefont {Sit}, \citenamefont {Zhang}, \citenamefont
  {Fickler}, \citenamefont {Miatto}, \citenamefont {Yao}, \citenamefont
  {Sciarrino},\ and\ \citenamefont {Karimi}}]{Bouchard:2021}%
  \BibitemOpen
  \bibfield  {author} {\bibinfo {author} {\bibfnamefont {F.}~\bibnamefont
  {Bouchard}}, \bibinfo {author} {\bibfnamefont {A.}~\bibnamefont {Sit}},
  \bibinfo {author} {\bibfnamefont {Y.}~\bibnamefont {Zhang}}, \bibinfo
  {author} {\bibfnamefont {R.}~\bibnamefont {Fickler}}, \bibinfo {author}
  {\bibfnamefont {F.~M.}\ \bibnamefont {Miatto}}, \bibinfo {author}
  {\bibfnamefont {Y.}~\bibnamefont {Yao}}, \bibinfo {author} {\bibfnamefont
  {F.}~\bibnamefont {Sciarrino}}, \ and\ \bibinfo {author} {\bibfnamefont
  {E.}~\bibnamefont {Karimi}},\ }\bibfield  {title} {\enquote {\bibinfo {title}
  {Two-photon interference: the hong–ou–mandel effect},}\ }\href {\doibase
  10.1088/1361-6633/abcd7a} {\bibfield  {journal} {\bibinfo  {journal} {Rep.
  Prog. Phys.}\ }\textbf {\bibinfo {volume} {84}},\ \bibinfo {pages} {012402}
  (\bibinfo {year} {2021})}\BibitemShut {NoStop}%
\bibitem [{\citenamefont {Mittal}\ \emph {et~al.}(2021)\citenamefont {Mittal},
  \citenamefont {Orre}, \citenamefont {Goldschmidt},\ and\ \citenamefont
  {Hafezi}}]{Mittal:2021}%
  \BibitemOpen
  \bibfield  {author} {\bibinfo {author} {\bibfnamefont {S.}~\bibnamefont
  {Mittal}}, \bibinfo {author} {\bibfnamefont {V.~V.}\ \bibnamefont {Orre}},
  \bibinfo {author} {\bibfnamefont {E.~A.}\ \bibnamefont {Goldschmidt}}, \ and\
  \bibinfo {author} {\bibfnamefont {M.}~\bibnamefont {Hafezi}},\ }\bibfield
  {title} {\enquote {\bibinfo {title} {Tunable quantum interference using a
  topological source of indistinguishable photon pairs},}\ }\href {\doibase
  10.1038/s41566-021-00810-1} {\bibfield  {journal} {\bibinfo  {journal}
  {Nature Photonics}\ }\textbf {\bibinfo {volume} {15}},\ \bibinfo {pages}
  {542--548} (\bibinfo {year} {2021})}\BibitemShut {NoStop}%
\bibitem [{\citenamefont {S{\"o}llner}\ \emph {et~al.}(2015)\citenamefont
  {S{\"o}llner}, \citenamefont {Mahmoodian}, \citenamefont {Hansen},
  \citenamefont {Midolo}, \citenamefont {Javadi}, \citenamefont
  {Kir{\v{s}}ansk{\.{e}}}, \citenamefont {Pregnolato}, \citenamefont {El-Ella},
  \citenamefont {Lee}, \citenamefont {Song}, \citenamefont {Stobbe},\ and\
  \citenamefont {Lodahl}}]{Lodahl:2015}%
  \BibitemOpen
  \bibfield  {author} {\bibinfo {author} {\bibfnamefont {I.}~\bibnamefont
  {S{\"o}llner}}, \bibinfo {author} {\bibfnamefont {S.}~\bibnamefont
  {Mahmoodian}}, \bibinfo {author} {\bibfnamefont {S.~L.}\ \bibnamefont
  {Hansen}}, \bibinfo {author} {\bibfnamefont {L.}~\bibnamefont {Midolo}},
  \bibinfo {author} {\bibfnamefont {A.}~\bibnamefont {Javadi}}, \bibinfo
  {author} {\bibfnamefont {G.}~\bibnamefont {Kir{\v{s}}ansk{\.{e}}}}, \bibinfo
  {author} {\bibfnamefont {T.}~\bibnamefont {Pregnolato}}, \bibinfo {author}
  {\bibfnamefont {H.}~\bibnamefont {El-Ella}}, \bibinfo {author} {\bibfnamefont
  {E.~H.}\ \bibnamefont {Lee}}, \bibinfo {author} {\bibfnamefont {J.~D.}\
  \bibnamefont {Song}}, \bibinfo {author} {\bibfnamefont {S.}~\bibnamefont
  {Stobbe}}, \ and\ \bibinfo {author} {\bibfnamefont {P.}~\bibnamefont
  {Lodahl}},\ }\bibfield  {title} {\enquote {\bibinfo {title} {Deterministic
  photon--emitter coupling in chiral photonic circuits},}\ }\href {\doibase
  10.1038/nnano.2015.159} {\bibfield  {journal} {\bibinfo  {journal} {Nature
  Nanotechnology}\ }\textbf {\bibinfo {volume} {10}},\ \bibinfo {pages}
  {775--778} (\bibinfo {year} {2015})}\BibitemShut {NoStop}%
\bibitem [{\citenamefont {Lodahl}, \citenamefont {Mahmoodian},\ and\
  \citenamefont {Stobbe}(2015)}]{Lodahl_Review:2015}%
  \BibitemOpen
  \bibfield  {author} {\bibinfo {author} {\bibfnamefont {P.}~\bibnamefont
  {Lodahl}}, \bibinfo {author} {\bibfnamefont {S.}~\bibnamefont {Mahmoodian}},
  \ and\ \bibinfo {author} {\bibfnamefont {S.}~\bibnamefont {Stobbe}},\
  }\bibfield  {title} {\enquote {\bibinfo {title} {Interfacing single photons
  and single quantum dots with photonic nanostructures},}\ }\href {\doibase
  10.1103/RevModPhys.87.347} {\bibfield  {journal} {\bibinfo  {journal} {Rev.
  Mod. Phys.}\ }\textbf {\bibinfo {volume} {87}},\ \bibinfo {pages} {347--400}
  (\bibinfo {year} {2015})}\BibitemShut {NoStop}%
\bibitem [{\citenamefont {Rodriguez-Fortuno}\ \emph {et~al.}(2013)\citenamefont
  {Rodriguez-Fortuno}, \citenamefont {Marino}, \citenamefont {Ginzburg},
  \citenamefont {O{\textquotesingle}Connor}, \citenamefont {Martinez},
  \citenamefont {Wurtz},\ and\ \citenamefont {Zayats}}]{RodriguezFortuno:2013}%
  \BibitemOpen
  \bibfield  {author} {\bibinfo {author} {\bibfnamefont {F.~J.}\ \bibnamefont
  {Rodriguez-Fortuno}}, \bibinfo {author} {\bibfnamefont {G.}~\bibnamefont
  {Marino}}, \bibinfo {author} {\bibfnamefont {P.}~\bibnamefont {Ginzburg}},
  \bibinfo {author} {\bibfnamefont {D.}~\bibnamefont
  {O{\textquotesingle}Connor}}, \bibinfo {author} {\bibfnamefont
  {A.}~\bibnamefont {Martinez}}, \bibinfo {author} {\bibfnamefont {G.~A.}\
  \bibnamefont {Wurtz}}, \ and\ \bibinfo {author} {\bibfnamefont {A.~V.}\
  \bibnamefont {Zayats}},\ }\bibfield  {title} {\enquote {\bibinfo {title}
  {Near-field interference for the unidirectional excitation of electromagnetic
  guided modes},}\ }\href {\doibase 10.1126/science.1233739} {\bibfield
  {journal} {\bibinfo  {journal} {Science}\ }\textbf {\bibinfo {volume}
  {340}},\ \bibinfo {pages} {328--330} (\bibinfo {year} {2013})}\BibitemShut
  {NoStop}%
\bibitem [{\citenamefont {Ozawa}\ \emph {et~al.}(2019)\citenamefont {Ozawa},
  \citenamefont {Price}, \citenamefont {Amo}, \citenamefont {Goldman},
  \citenamefont {Hafezi}, \citenamefont {Lu}, \citenamefont {Rechtsman},
  \citenamefont {Schuster}, \citenamefont {Simon}, \citenamefont {Zilberberg},\
  and\ \citenamefont {Carusotto}}]{Ozawa_Review:2019}%
  \BibitemOpen
  \bibfield  {author} {\bibinfo {author} {\bibfnamefont {T.}~\bibnamefont
  {Ozawa}}, \bibinfo {author} {\bibfnamefont {H.~M.}\ \bibnamefont {Price}},
  \bibinfo {author} {\bibfnamefont {A.}~\bibnamefont {Amo}}, \bibinfo {author}
  {\bibfnamefont {N.}~\bibnamefont {Goldman}}, \bibinfo {author} {\bibfnamefont
  {M.}~\bibnamefont {Hafezi}}, \bibinfo {author} {\bibfnamefont
  {L.}~\bibnamefont {Lu}}, \bibinfo {author} {\bibfnamefont {M.~C.}\
  \bibnamefont {Rechtsman}}, \bibinfo {author} {\bibfnamefont {D.}~\bibnamefont
  {Schuster}}, \bibinfo {author} {\bibfnamefont {J.}~\bibnamefont {Simon}},
  \bibinfo {author} {\bibfnamefont {O.}~\bibnamefont {Zilberberg}}, \ and\
  \bibinfo {author} {\bibfnamefont {I.}~\bibnamefont {Carusotto}},\ }\bibfield
  {title} {\enquote {\bibinfo {title} {Topological photonics},}\ }\href
  {\doibase 10.1103/RevModPhys.91.015006} {\bibfield  {journal} {\bibinfo
  {journal} {Rev. Mod. Phys.}\ }\textbf {\bibinfo {volume} {91}},\ \bibinfo
  {pages} {015006} (\bibinfo {year} {2019})}\BibitemShut {NoStop}%
\bibitem [{\citenamefont {Khanikaev}\ and\ \citenamefont
  {Shvets}(2017)}]{Shvets_Review:2017}%
  \BibitemOpen
  \bibfield  {author} {\bibinfo {author} {\bibfnamefont {A.~B.}\ \bibnamefont
  {Khanikaev}}\ and\ \bibinfo {author} {\bibfnamefont {G.}~\bibnamefont
  {Shvets}},\ }\bibfield  {title} {\enquote {\bibinfo {title} {Two-dimensional
  topological photonics},}\ }\href {\doibase 10.1038/s41566-017-0048-5}
  {\bibfield  {journal} {\bibinfo  {journal} {Nature Photonics}\ }\textbf
  {\bibinfo {volume} {11}},\ \bibinfo {pages} {763--773} (\bibinfo {year}
  {2017})}\BibitemShut {NoStop}%
\bibitem [{\citenamefont {Lu}, \citenamefont {Joannopoulos},\ and\
  \citenamefont {Solja\ifmmode \check{c}\else
  \v{c}\fi{}i\ifmmode~\acute{c}\else \'{c}\fi{}}(2014)}]{Lu_Review:2014}%
  \BibitemOpen
  \bibfield  {author} {\bibinfo {author} {\bibfnamefont {L.}~\bibnamefont
  {Lu}}, \bibinfo {author} {\bibfnamefont {J.~D.}\ \bibnamefont
  {Joannopoulos}}, \ and\ \bibinfo {author} {\bibfnamefont {M.}~\bibnamefont
  {Solja\ifmmode \check{c}\else \v{c}\fi{}i\ifmmode~\acute{c}\else
  \'{c}\fi{}}},\ }\bibfield  {title} {\enquote {\bibinfo {title} {Topological
  photonics},}\ }\href {https://doi.org/10.1038/nphoton.2014.248} {\bibfield
  {journal} {\bibinfo  {journal} {Nature Photonics}\ }\textbf {\bibinfo
  {volume} {8}},\ \bibinfo {pages} {821--829} (\bibinfo {year}
  {2014})}\BibitemShut {NoStop}%
\bibitem [{\citenamefont {Haldane}\ and\ \citenamefont
  {Raghu}(2008)}]{Raghu:2008_1}%
  \BibitemOpen
  \bibfield  {author} {\bibinfo {author} {\bibfnamefont {F.~D.~M.}\
  \bibnamefont {Haldane}}\ and\ \bibinfo {author} {\bibfnamefont
  {S.}~\bibnamefont {Raghu}},\ }\bibfield  {title} {\enquote {\bibinfo {title}
  {Possible realization of directional optical waveguides in photonic crystals
  with broken time-reversal symmetry},}\ }\href {\doibase
  10.1103/PhysRevLett.100.013904} {\bibfield  {journal} {\bibinfo  {journal}
  {Phys. Rev. Lett.}\ }\textbf {\bibinfo {volume} {100}},\ \bibinfo {pages}
  {013904} (\bibinfo {year} {2008})}\BibitemShut {NoStop}%
\bibitem [{\citenamefont {Wang}\ \emph {et~al.}(2009)\citenamefont {Wang},
  \citenamefont {Chong}, \citenamefont {Joannopoulos},\ and\ \citenamefont
  {Solja\ifmmode \check{c}\else \v{c}\fi{}i\ifmmode~\acute{c}\else
  \'{c}\fi{}}}]{WangZheng:2009}%
  \BibitemOpen
  \bibfield  {author} {\bibinfo {author} {\bibfnamefont {Z.}~\bibnamefont
  {Wang}}, \bibinfo {author} {\bibfnamefont {Y.~D.}\ \bibnamefont {Chong}},
  \bibinfo {author} {\bibfnamefont {J.~D.}\ \bibnamefont {Joannopoulos}}, \
  and\ \bibinfo {author} {\bibfnamefont {M.}~\bibnamefont {Solja\ifmmode
  \check{c}\else \v{c}\fi{}i\ifmmode~\acute{c}\else \'{c}\fi{}}},\ }\bibfield
  {title} {\enquote {\bibinfo {title} {Observation of unidirectional
  backscattering-immune topological electromagnetic states},}\ }\href {\doibase
  10.1038/nature08293} {\bibfield  {journal} {\bibinfo  {journal} {Nature}\
  }\textbf {\bibinfo {volume} {461}},\ \bibinfo {pages} {772--775} (\bibinfo
  {year} {2009})}\BibitemShut {NoStop}%
\bibitem [{\citenamefont {Khanikaev}\ \emph {et~al.}(2012)\citenamefont
  {Khanikaev}, \citenamefont {Hossein~Mousavi}, \citenamefont {Tse},
  \citenamefont {Kargarian}, \citenamefont {MacDonald},\ and\ \citenamefont
  {Shvets}}]{Khanikaev:2012}%
  \BibitemOpen
  \bibfield  {author} {\bibinfo {author} {\bibfnamefont {A.~B.}\ \bibnamefont
  {Khanikaev}}, \bibinfo {author} {\bibfnamefont {S.}~\bibnamefont
  {Hossein~Mousavi}}, \bibinfo {author} {\bibfnamefont {W.-K.}\ \bibnamefont
  {Tse}}, \bibinfo {author} {\bibfnamefont {M.}~\bibnamefont {Kargarian}},
  \bibinfo {author} {\bibfnamefont {A.~H.}\ \bibnamefont {MacDonald}}, \ and\
  \bibinfo {author} {\bibfnamefont {G.}~\bibnamefont {Shvets}},\ }\bibfield
  {title} {\enquote {\bibinfo {title} {Photonic topological insulators},}\
  }\href {https://doi.org/10.1038/nmat3520} {\bibfield  {journal} {\bibinfo
  {journal} {Nature Materials}\ }\textbf {\bibinfo {volume} {12}},\ \bibinfo
  {pages} {233--239} (\bibinfo {year} {2012})}\BibitemShut {NoStop}%
\bibitem [{\citenamefont {Chen}\ \emph {et~al.}(2014)\citenamefont {Chen},
  \citenamefont {Jiang}, \citenamefont {Chen}, \citenamefont {Zhu},
  \citenamefont {Zhou}, \citenamefont {Dong},\ and\ \citenamefont
  {Chan}}]{CTChan:2014}%
  \BibitemOpen
  \bibfield  {author} {\bibinfo {author} {\bibfnamefont {W.-J.}\ \bibnamefont
  {Chen}}, \bibinfo {author} {\bibfnamefont {S.-J.}\ \bibnamefont {Jiang}},
  \bibinfo {author} {\bibfnamefont {X.-D.}\ \bibnamefont {Chen}}, \bibinfo
  {author} {\bibfnamefont {B.}~\bibnamefont {Zhu}}, \bibinfo {author}
  {\bibfnamefont {L.}~\bibnamefont {Zhou}}, \bibinfo {author} {\bibfnamefont
  {J.-W.}\ \bibnamefont {Dong}}, \ and\ \bibinfo {author} {\bibfnamefont
  {C.~T.}\ \bibnamefont {Chan}},\ }\bibfield  {title} {\enquote {\bibinfo
  {title} {Experimental realization of photonic topological insulator in a
  uniaxial metacrystal waveguide},}\ }\href {\doibase 10.1038/ncomms6782}
  {\bibfield  {journal} {\bibinfo  {journal} {Nature Communications}\ }\textbf
  {\bibinfo {volume} {5}},\ \bibinfo {pages} {5782} (\bibinfo {year}
  {2014})}\BibitemShut {NoStop}%
\bibitem [{\citenamefont {Ma}\ \emph {et~al.}(2015)\citenamefont {Ma},
  \citenamefont {Khanikaev}, \citenamefont {Mousavi},\ and\ \citenamefont
  {Shvets}}]{Ma:2015}%
  \BibitemOpen
  \bibfield  {author} {\bibinfo {author} {\bibfnamefont {T.}~\bibnamefont
  {Ma}}, \bibinfo {author} {\bibfnamefont {A.~B.}\ \bibnamefont {Khanikaev}},
  \bibinfo {author} {\bibfnamefont {S.~H.}\ \bibnamefont {Mousavi}}, \ and\
  \bibinfo {author} {\bibfnamefont {G.}~\bibnamefont {Shvets}},\ }\bibfield
  {title} {\enquote {\bibinfo {title} {Guiding electromagnetic waves around
  sharp corners: Topologically protected photonic transport in
  metawaveguides},}\ }\href {\doibase 10.1103/PhysRevLett.114.127401}
  {\bibfield  {journal} {\bibinfo  {journal} {Phys. Rev. Lett.}\ }\textbf
  {\bibinfo {volume} {114}},\ \bibinfo {pages} {127401} (\bibinfo {year}
  {2015})}\BibitemShut {NoStop}%
\bibitem [{\citenamefont {Lai}\ \emph {et~al.}(2016)\citenamefont {Lai},
  \citenamefont {Ma}, \citenamefont {Bo}, \citenamefont {Anlage},\ and\
  \citenamefont {Shvets}}]{Lai:2016}%
  \BibitemOpen
  \bibfield  {author} {\bibinfo {author} {\bibfnamefont {K.}~\bibnamefont
  {Lai}}, \bibinfo {author} {\bibfnamefont {T.}~\bibnamefont {Ma}}, \bibinfo
  {author} {\bibfnamefont {X.}~\bibnamefont {Bo}}, \bibinfo {author}
  {\bibfnamefont {S.}~\bibnamefont {Anlage}}, \ and\ \bibinfo {author}
  {\bibfnamefont {G.}~\bibnamefont {Shvets}},\ }\bibfield  {title} {\enquote
  {\bibinfo {title} {Experimental realization of a reflections-free compact
  delay line based on a photonic topological insulator},}\ }\href
  {https://doi.org/10.1038/srep28453} {\bibfield  {journal} {\bibinfo
  {journal} {Scientific Reports}\ }\textbf {\bibinfo {volume} {6}},\ \bibinfo
  {pages} {28453} (\bibinfo {year} {2016})}\BibitemShut {NoStop}%
\bibitem [{\citenamefont {Ma}\ and\ \citenamefont {Shvets}(2017)}]{Ma:2017}%
  \BibitemOpen
  \bibfield  {author} {\bibinfo {author} {\bibfnamefont {T.}~\bibnamefont
  {Ma}}\ and\ \bibinfo {author} {\bibfnamefont {G.}~\bibnamefont {Shvets}},\
  }\bibfield  {title} {\enquote {\bibinfo {title} {Scattering-free edge states
  between heterogeneous photonic topological insulators},}\ }\href {\doibase
  10.1103/PhysRevB.95.165102} {\bibfield  {journal} {\bibinfo  {journal} {Phys.
  Rev. B}\ }\textbf {\bibinfo {volume} {95}},\ \bibinfo {pages} {165102}
  (\bibinfo {year} {2017})}\BibitemShut {NoStop}%
\bibitem [{\citenamefont {Ma}\ and\ \citenamefont {Shvets}(2016)}]{Ma:2016}%
  \BibitemOpen
  \bibfield  {author} {\bibinfo {author} {\bibfnamefont {T.}~\bibnamefont
  {Ma}}\ and\ \bibinfo {author} {\bibfnamefont {G.}~\bibnamefont {Shvets}},\
  }\bibfield  {title} {\enquote {\bibinfo {title} {All-si valley-hall photonic
  topological insulator},}\ }\href {\doibase 10.1088/1367-2630/18/2/025012}
  {\bibfield  {journal} {\bibinfo  {journal} {New Journal of Physics}\ }\textbf
  {\bibinfo {volume} {18}},\ \bibinfo {pages} {025012} (\bibinfo {year}
  {2016})}\BibitemShut {NoStop}%
\bibitem [{\citenamefont {Gao}\ \emph {et~al.}(2018)\citenamefont {Gao},
  \citenamefont {Xue}, \citenamefont {Yang}, \citenamefont {Lai}, \citenamefont
  {Yu}, \citenamefont {Lin}, \citenamefont {Chong}, \citenamefont {Shvets},\
  and\ \citenamefont {Zhang}}]{GaoFei:2018}%
  \BibitemOpen
  \bibfield  {author} {\bibinfo {author} {\bibfnamefont {F.}~\bibnamefont
  {Gao}}, \bibinfo {author} {\bibfnamefont {H.}~\bibnamefont {Xue}}, \bibinfo
  {author} {\bibfnamefont {Z.}~\bibnamefont {Yang}}, \bibinfo {author}
  {\bibfnamefont {K.}~\bibnamefont {Lai}}, \bibinfo {author} {\bibfnamefont
  {Y.}~\bibnamefont {Yu}}, \bibinfo {author} {\bibfnamefont {X.}~\bibnamefont
  {Lin}}, \bibinfo {author} {\bibfnamefont {Y.}~\bibnamefont {Chong}}, \bibinfo
  {author} {\bibfnamefont {G.}~\bibnamefont {Shvets}}, \ and\ \bibinfo {author}
  {\bibfnamefont {B.}~\bibnamefont {Zhang}},\ }\bibfield  {title} {\enquote
  {\bibinfo {title} {Topologically protected refraction of robust kink states
  in valley photonic crystals},}\ }\href {\doibase 10.1038/nphys4304}
  {\bibfield  {journal} {\bibinfo  {journal} {Nature Physics}\ }\textbf
  {\bibinfo {volume} {14}},\ \bibinfo {pages} {140--144} (\bibinfo {year}
  {2018})}\BibitemShut {NoStop}%
\bibitem [{\citenamefont {Shalaev}\ \emph {et~al.}(2019)\citenamefont
  {Shalaev}, \citenamefont {Walasik}, \citenamefont {Tsukernik}, \citenamefont
  {Xu},\ and\ \citenamefont {Litchinitser}}]{Litchinitser:2019}%
  \BibitemOpen
  \bibfield  {author} {\bibinfo {author} {\bibfnamefont {M.~I.}\ \bibnamefont
  {Shalaev}}, \bibinfo {author} {\bibfnamefont {W.}~\bibnamefont {Walasik}},
  \bibinfo {author} {\bibfnamefont {A.}~\bibnamefont {Tsukernik}}, \bibinfo
  {author} {\bibfnamefont {Y.}~\bibnamefont {Xu}}, \ and\ \bibinfo {author}
  {\bibfnamefont {N.~M.}\ \bibnamefont {Litchinitser}},\ }\bibfield  {title}
  {\enquote {\bibinfo {title} {Robust topologically protected transport in
  photonic crystals at telecommunication wavelengths},}\ }\href {\doibase
  10.1038/s41565-018-0297-6} {\bibfield  {journal} {\bibinfo  {journal} {Nature
  Nanotechnology}\ }\textbf {\bibinfo {volume} {14}},\ \bibinfo {pages}
  {31--34} (\bibinfo {year} {2019})}\BibitemShut {NoStop}%
\bibitem [{\citenamefont {Dong}\ \emph {et~al.}(2017)\citenamefont {Dong},
  \citenamefont {Chen}, \citenamefont {Zhu}, \citenamefont {Wang},\ and\
  \citenamefont {Zhang}}]{ZhangXiang:2017}%
  \BibitemOpen
  \bibfield  {author} {\bibinfo {author} {\bibfnamefont {J.-W.}\ \bibnamefont
  {Dong}}, \bibinfo {author} {\bibfnamefont {X.-D.}\ \bibnamefont {Chen}},
  \bibinfo {author} {\bibfnamefont {H.}~\bibnamefont {Zhu}}, \bibinfo {author}
  {\bibfnamefont {Y.}~\bibnamefont {Wang}}, \ and\ \bibinfo {author}
  {\bibfnamefont {X.}~\bibnamefont {Zhang}},\ }\bibfield  {title} {\enquote
  {\bibinfo {title} {Valley photonic crystals for control of spin
  and topology},}\ }\href {\doibase 10.1038/nmat4807} {\bibfield  {journal}
  {\bibinfo  {journal} {Nature Materials}\ }\textbf {\bibinfo {volume} {16}},\
  \bibinfo {pages} {298--302} (\bibinfo {year} {2017})}\BibitemShut {NoStop}%
\bibitem [{\citenamefont {Noh}\ \emph {et~al.}(2018)\citenamefont {Noh},
  \citenamefont {Huang}, \citenamefont {Chen},\ and\ \citenamefont
  {Rechtsman}}]{Rechtsman:2018}%
  \BibitemOpen
  \bibfield  {author} {\bibinfo {author} {\bibfnamefont {J.}~\bibnamefont
  {Noh}}, \bibinfo {author} {\bibfnamefont {S.}~\bibnamefont {Huang}}, \bibinfo
  {author} {\bibfnamefont {K.~P.}\ \bibnamefont {Chen}}, \ and\ \bibinfo
  {author} {\bibfnamefont {M.~C.}\ \bibnamefont {Rechtsman}},\ }\bibfield
  {title} {\enquote {\bibinfo {title} {Observation of photonic topological
  valley hall edge states},}\ }\href {\doibase 10.1103/PhysRevLett.120.063902}
  {\bibfield  {journal} {\bibinfo  {journal} {Phys. Rev. Lett.}\ }\textbf
  {\bibinfo {volume} {120}},\ \bibinfo {pages} {063902} (\bibinfo {year}
  {2018})}\BibitemShut {NoStop}%
\bibitem [{\citenamefont {Ma}\ and\ \citenamefont
  {Anlage}(2020)}]{Anlage_Review:2020}%
  \BibitemOpen
  \bibfield  {author} {\bibinfo {author} {\bibfnamefont {S.}~\bibnamefont
  {Ma}}\ and\ \bibinfo {author} {\bibfnamefont {S.~M.}\ \bibnamefont
  {Anlage}},\ }\bibfield  {title} {\enquote {\bibinfo {title} {Microwave
  applications of photonic topological insulators},}\ }\href {\doibase
  10.1063/5.0008046} {\bibfield  {journal} {\bibinfo  {journal} {Applied
  Physics Letters}\ }\textbf {\bibinfo {volume} {116}},\ \bibinfo {pages}
  {250502} (\bibinfo {year} {2020})}\BibitemShut {NoStop}%
\bibitem [{\citenamefont {Jin}\ \emph {et~al.}(2019)\citenamefont {Jin},
  \citenamefont {Yin}, \citenamefont {Ni}, \citenamefont
  {Solja{\v{c}}i{\'{c}}}, \citenamefont {Zhen},\ and\ \citenamefont
  {Peng}}]{PengChao:2019}%
  \BibitemOpen
  \bibfield  {author} {\bibinfo {author} {\bibfnamefont {J.}~\bibnamefont
  {Jin}}, \bibinfo {author} {\bibfnamefont {X.}~\bibnamefont {Yin}}, \bibinfo
  {author} {\bibfnamefont {L.}~\bibnamefont {Ni}}, \bibinfo {author}
  {\bibfnamefont {M.}~\bibnamefont {Solja{\v{c}}i{\'{c}}}}, \bibinfo {author}
  {\bibfnamefont {B.}~\bibnamefont {Zhen}}, \ and\ \bibinfo {author}
  {\bibfnamefont {C.}~\bibnamefont {Peng}},\ }\bibfield  {title} {\enquote
  {\bibinfo {title} {Topologically enabled ultrahigh-q guided resonances robust
  to out-of-plane scattering},}\ }\href {\doibase 10.1038/s41586-019-1664-7}
  {\bibfield  {journal} {\bibinfo  {journal} {Nature}\ }\textbf {\bibinfo
  {volume} {574}},\ \bibinfo {pages} {501--504} (\bibinfo {year}
  {2019})}\BibitemShut {NoStop}%
\bibitem [{\citenamefont {Ota}\ \emph {et~al.}(2019)\citenamefont {Ota},
  \citenamefont {Liu}, \citenamefont {Katsumi}, \citenamefont {Watanabe},
  \citenamefont {Wakabayashi}, \citenamefont {Arakawa},\ and\ \citenamefont
  {Iwamoto}}]{Ota:19}%
  \BibitemOpen
  \bibfield  {author} {\bibinfo {author} {\bibfnamefont {Y.}~\bibnamefont
  {Ota}}, \bibinfo {author} {\bibfnamefont {F.}~\bibnamefont {Liu}}, \bibinfo
  {author} {\bibfnamefont {R.}~\bibnamefont {Katsumi}}, \bibinfo {author}
  {\bibfnamefont {K.}~\bibnamefont {Watanabe}}, \bibinfo {author}
  {\bibfnamefont {K.}~\bibnamefont {Wakabayashi}}, \bibinfo {author}
  {\bibfnamefont {Y.}~\bibnamefont {Arakawa}}, \ and\ \bibinfo {author}
  {\bibfnamefont {S.}~\bibnamefont {Iwamoto}},\ }\bibfield  {title} {\enquote
  {\bibinfo {title} {Photonic crystal nanocavity based on a topological corner
  state},}\ }\href {\doibase 10.1364/OPTICA.6.000786} {\bibfield  {journal}
  {\bibinfo  {journal} {Optica}\ }\textbf {\bibinfo {volume} {6}},\ \bibinfo
  {pages} {786--789} (\bibinfo {year} {2019})}\BibitemShut {NoStop}%
\bibitem [{\citenamefont {Gao}\ \emph {et~al.}(2020)\citenamefont {Gao},
  \citenamefont {Yang}, \citenamefont {Lin}, \citenamefont {Zhang},
  \citenamefont {Li}, \citenamefont {Bo}, \citenamefont {Wang},\ and\
  \citenamefont {Lu}}]{LuLing:2019}%
  \BibitemOpen
  \bibfield  {author} {\bibinfo {author} {\bibfnamefont {X.}~\bibnamefont
  {Gao}}, \bibinfo {author} {\bibfnamefont {L.}~\bibnamefont {Yang}}, \bibinfo
  {author} {\bibfnamefont {H.}~\bibnamefont {Lin}}, \bibinfo {author}
  {\bibfnamefont {L.}~\bibnamefont {Zhang}}, \bibinfo {author} {\bibfnamefont
  {J.}~\bibnamefont {Li}}, \bibinfo {author} {\bibfnamefont {F.}~\bibnamefont
  {Bo}}, \bibinfo {author} {\bibfnamefont {Z.}~\bibnamefont {Wang}}, \ and\
  \bibinfo {author} {\bibfnamefont {L.}~\bibnamefont {Lu}},\ }\bibfield
  {title} {\enquote {\bibinfo {title} {Dirac-vortex topological cavities},}\
  }\href {\doibase 10.1038/s41565-020-0773-7} {\bibfield  {journal} {\bibinfo
  {journal} {Nature Nanotechnology}\ }\textbf {\bibinfo {volume} {15}},\
  \bibinfo {pages} {1012--1018} (\bibinfo {year} {2020})}\BibitemShut {NoStop}%
\bibitem [{\citenamefont {Li}\ \emph {et~al.}(2020)\citenamefont {Li},
  \citenamefont {Yu}, \citenamefont {Liu}, \citenamefont {Zhang},\ and\
  \citenamefont {Shvets}}]{Yandong:2020}%
  \BibitemOpen
  \bibfield  {author} {\bibinfo {author} {\bibfnamefont {Y.}~\bibnamefont
  {Li}}, \bibinfo {author} {\bibfnamefont {Y.}~\bibnamefont {Yu}}, \bibinfo
  {author} {\bibfnamefont {F.}~\bibnamefont {Liu}}, \bibinfo {author}
  {\bibfnamefont {B.}~\bibnamefont {Zhang}}, \ and\ \bibinfo {author}
  {\bibfnamefont {G.}~\bibnamefont {Shvets}},\ }\bibfield  {title} {\enquote
  {\bibinfo {title} {Topology-controlled photonic cavity based on the
  near-conservation of the valley degree of freedom},}\ }\href {\doibase
  10.1103/PhysRevLett.125.213902} {\bibfield  {journal} {\bibinfo  {journal}
  {Phys. Rev. Lett.}\ }\textbf {\bibinfo {volume} {125}},\ \bibinfo {pages}
  {213902} (\bibinfo {year} {2020})}\BibitemShut {NoStop}%
\bibitem [{\citenamefont {Harari}\ \emph {et~al.}(2018)\citenamefont {Harari},
  \citenamefont {Bandres}, \citenamefont {Lumer}, \citenamefont {Rechtsman},
  \citenamefont {Chong}, \citenamefont {Khajavikhan}, \citenamefont
  {Christodoulides},\ and\ \citenamefont {Segev}}]{TIlaser:Theory}%
  \BibitemOpen
  \bibfield  {author} {\bibinfo {author} {\bibfnamefont {G.}~\bibnamefont
  {Harari}}, \bibinfo {author} {\bibfnamefont {M.~A.}\ \bibnamefont {Bandres}},
  \bibinfo {author} {\bibfnamefont {Y.}~\bibnamefont {Lumer}}, \bibinfo
  {author} {\bibfnamefont {M.~C.}\ \bibnamefont {Rechtsman}}, \bibinfo {author}
  {\bibfnamefont {Y.~D.}\ \bibnamefont {Chong}}, \bibinfo {author}
  {\bibfnamefont {M.}~\bibnamefont {Khajavikhan}}, \bibinfo {author}
  {\bibfnamefont {D.~N.}\ \bibnamefont {Christodoulides}}, \ and\ \bibinfo
  {author} {\bibfnamefont {M.}~\bibnamefont {Segev}},\ }\bibfield  {title}
  {\enquote {\bibinfo {title} {Topological insulator laser: Theory},}\ }\href
  {\doibase 10.1126/science.aar4003} {\bibfield  {journal} {\bibinfo  {journal}
  {Science}\ }\textbf {\bibinfo {volume} {359}} (\bibinfo {year} {2018}),\
  10.1126/science.aar4003}\BibitemShut {NoStop}%
\bibitem [{\citenamefont {Bandres}\ \emph {et~al.}(2018)\citenamefont
  {Bandres}, \citenamefont {Wittek}, \citenamefont {Harari}, \citenamefont
  {Parto}, \citenamefont {Ren}, \citenamefont {Segev}, \citenamefont
  {Christodoulides},\ and\ \citenamefont {Khajavikhan}}]{TIlaser:Experiment}%
  \BibitemOpen
  \bibfield  {author} {\bibinfo {author} {\bibfnamefont {M.~A.}\ \bibnamefont
  {Bandres}}, \bibinfo {author} {\bibfnamefont {S.}~\bibnamefont {Wittek}},
  \bibinfo {author} {\bibfnamefont {G.}~\bibnamefont {Harari}}, \bibinfo
  {author} {\bibfnamefont {M.}~\bibnamefont {Parto}}, \bibinfo {author}
  {\bibfnamefont {J.}~\bibnamefont {Ren}}, \bibinfo {author} {\bibfnamefont
  {M.}~\bibnamefont {Segev}}, \bibinfo {author} {\bibfnamefont {D.~N.}\
  \bibnamefont {Christodoulides}}, \ and\ \bibinfo {author} {\bibfnamefont
  {M.}~\bibnamefont {Khajavikhan}},\ }\bibfield  {title} {\enquote {\bibinfo
  {title} {Topological insulator laser: Experiments},}\ }\href {\doibase
  10.1126/science.aar4005} {\bibfield  {journal} {\bibinfo  {journal}
  {Science}\ }\textbf {\bibinfo {volume} {359}} (\bibinfo {year} {2018}),\
  10.1126/science.aar4005}\BibitemShut {NoStop}%
\bibitem [{\citenamefont {Zeng}\ \emph {et~al.}(2020)\citenamefont {Zeng},
  \citenamefont {Chattopadhyay}, \citenamefont {Zhu}, \citenamefont {Qiang},
  \citenamefont {Li}, \citenamefont {Jin}, \citenamefont {Li}, \citenamefont
  {Davies}, \citenamefont {Linfield}, \citenamefont {Zhang}, \citenamefont
  {Chong},\ and\ \citenamefont {Wang}}]{ZhangBaile:2020}%
  \BibitemOpen
  \bibfield  {author} {\bibinfo {author} {\bibfnamefont {Y.}~\bibnamefont
  {Zeng}}, \bibinfo {author} {\bibfnamefont {U.}~\bibnamefont {Chattopadhyay}},
  \bibinfo {author} {\bibfnamefont {B.}~\bibnamefont {Zhu}}, \bibinfo {author}
  {\bibfnamefont {B.}~\bibnamefont {Qiang}}, \bibinfo {author} {\bibfnamefont
  {J.}~\bibnamefont {Li}}, \bibinfo {author} {\bibfnamefont {Y.}~\bibnamefont
  {Jin}}, \bibinfo {author} {\bibfnamefont {L.}~\bibnamefont {Li}}, \bibinfo
  {author} {\bibfnamefont {A.~G.}\ \bibnamefont {Davies}}, \bibinfo {author}
  {\bibfnamefont {E.~H.}\ \bibnamefont {Linfield}}, \bibinfo {author}
  {\bibfnamefont {B.}~\bibnamefont {Zhang}}, \bibinfo {author} {\bibfnamefont
  {Y.}~\bibnamefont {Chong}}, \ and\ \bibinfo {author} {\bibfnamefont {Q.~J.}\
  \bibnamefont {Wang}},\ }\bibfield  {title} {\enquote {\bibinfo {title}
  {Electrically pumped topological laser with valley edge modes},}\ }\href
  {\doibase 10.1038/s41586-020-1981-x} {\bibfield  {journal} {\bibinfo
  {journal} {Nature}\ }\textbf {\bibinfo {volume} {578}},\ \bibinfo {pages}
  {246--250} (\bibinfo {year} {2020})}\BibitemShut {NoStop}%
\bibitem [{\citenamefont {Cerf}\ and\ \citenamefont
  {Jabbour}(2020)}]{Cerf:2020}%
  \BibitemOpen
  \bibfield  {author} {\bibinfo {author} {\bibfnamefont {N.~J.}\ \bibnamefont
  {Cerf}}\ and\ \bibinfo {author} {\bibfnamefont {M.~G.}\ \bibnamefont
  {Jabbour}},\ }\bibfield  {title} {\enquote {\bibinfo {title} {Two-boson
  quantum interference in time},}\ }\href {\doibase 10.1073/pnas.2010827117}
  {\bibfield  {journal} {\bibinfo  {journal} {Proceedings of the National
  Academy of Sciences}\ }\textbf {\bibinfo {volume} {117}},\ \bibinfo {pages}
  {33107--33116} (\bibinfo {year} {2020})}\BibitemShut {NoStop}%
\bibitem [{\citenamefont {Sanaka}, \citenamefont {Resch},\ and\ \citenamefont
  {Zeilinger}(2006)}]{Zeilinger:2006}%
  \BibitemOpen
  \bibfield  {author} {\bibinfo {author} {\bibfnamefont {K.}~\bibnamefont
  {Sanaka}}, \bibinfo {author} {\bibfnamefont {K.~J.}\ \bibnamefont {Resch}}, \
  and\ \bibinfo {author} {\bibfnamefont {A.}~\bibnamefont {Zeilinger}},\
  }\bibfield  {title} {\enquote {\bibinfo {title} {Filtering out photonic fock
  states},}\ }\href {\doibase 10.1103/PhysRevLett.96.083601} {\bibfield
  {journal} {\bibinfo  {journal} {Phys. Rev. Lett.}\ }\textbf {\bibinfo
  {volume} {96}},\ \bibinfo {pages} {083601} (\bibinfo {year}
  {2006})}\BibitemShut {NoStop}%
\bibitem [{\citenamefont {Kang}\ \emph {et~al.}(2018)\citenamefont {Kang},
  \citenamefont {Ni}, \citenamefont {Cheng}, \citenamefont {Khanikaev},\ and\
  \citenamefont {Genack}}]{KangYuhao:2018}%
  \BibitemOpen
  \bibfield  {author} {\bibinfo {author} {\bibfnamefont {Y.}~\bibnamefont
  {Kang}}, \bibinfo {author} {\bibfnamefont {X.}~\bibnamefont {Ni}}, \bibinfo
  {author} {\bibfnamefont {X.}~\bibnamefont {Cheng}}, \bibinfo {author}
  {\bibfnamefont {A.~B.}\ \bibnamefont {Khanikaev}}, \ and\ \bibinfo {author}
  {\bibfnamefont {A.~Z.}\ \bibnamefont {Genack}},\ }\bibfield  {title}
  {\enquote {\bibinfo {title} {Pseudo-spin--valley coupled edge states in a
  photonic topological insulator},}\ }\href {\doibase
  10.1038/s41467-018-05408-w} {\bibfield  {journal} {\bibinfo  {journal}
  {Nature Communications}\ }\textbf {\bibinfo {volume} {9}},\ \bibinfo {pages}
  {3029} (\bibinfo {year} {2018})}\BibitemShut {NoStop}%
\bibitem [{\citenamefont {Bo}\ \emph {et~al.}(2016)\citenamefont {Bo},
  \citenamefont {Lai}, \citenamefont {Yu}, \citenamefont {Ma}, \citenamefont
  {Shvets},\ and\ \citenamefont {Anlage}}]{Anlage:2016}%
  \BibitemOpen
  \bibfield  {author} {\bibinfo {author} {\bibfnamefont {X.}~\bibnamefont
  {Bo}}, \bibinfo {author} {\bibfnamefont {K.}~\bibnamefont {Lai}}, \bibinfo
  {author} {\bibfnamefont {Y.}~\bibnamefont {Yu}}, \bibinfo {author}
  {\bibfnamefont {T.}~\bibnamefont {Ma}}, \bibinfo {author} {\bibfnamefont
  {G.}~\bibnamefont {Shvets}}, \ and\ \bibinfo {author} {\bibfnamefont
  {S.}~\bibnamefont {Anlage}},\ }\bibfield  {title} {\enquote {\bibinfo {title}
  {Exciting reflectionless unidirectional edge modes in a reciprocal photonic
  topological insulator medium},}\ }\href {\doibase 10.1103/PhysRevB.94.195427}
  {\bibfield  {journal} {\bibinfo  {journal} {Phys. Rev. B}\ }\textbf {\bibinfo
  {volume} {94}},\ \bibinfo {pages} {195427} (\bibinfo {year}
  {2016})}\BibitemShut {NoStop}%
\bibitem [{\citenamefont {F.~Picardi}, \citenamefont {V.~Zayats},\ and\
  \citenamefont {J.~Rodríguez-Fortuño}(2019)}]{Zayats:2019}%
  \BibitemOpen
  \bibfield  {author} {\bibinfo {author} {\bibfnamefont {M.}~\bibnamefont
  {F.~Picardi}}, \bibinfo {author} {\bibfnamefont {A.}~\bibnamefont
  {V.~Zayats}}, \ and\ \bibinfo {author} {\bibfnamefont {F.}~\bibnamefont
  {J.~Rodríguez-Fortuño}},\ }\bibfield  {title} {\enquote {\bibinfo {title}
  {Amplitude and phase control of guided modes excitation from a single dipole
  source: Engineering far- and near-field directionality},}\ }\href {\doibase
  https://doi.org/10.1002/lpor.201900250} {\bibfield  {journal} {\bibinfo
  {journal} {Laser \& Photonics Reviews}\ }\textbf {\bibinfo {volume} {13}},\
  \bibinfo {pages} {1900250} (\bibinfo {year} {2019})}\BibitemShut {NoStop}%
\end{thebibliography}

\begin{thebibliography}{1}

\bibitem{Kane_Mele:05}
C.~L. Kane and E.~J. Mele.
\newblock Quantum spin hall effect in graphene.
\newblock {\em Phys. Rev. Lett.}, 95:226801, Nov 2005.

\bibitem{MaTzuhsuan:15}
Tzuhsuan Ma, Alexander~B. Khanikaev, S.~Hossein Mousavi, and Gennady Shvets.
\newblock Guiding electromagnetic waves around sharp corners: Topologically
  protected photonic transport in metawaveguides.
\newblock {\em Phys. Rev. Lett.}, 114:127401, Mar 2015.

\bibitem{MaTzuhsuan:17}
Tzuhsuan Ma and Gennady Shvets.
\newblock Scattering-free edge states between heterogeneous photonic
  topological insulators.
\newblock {\em Phys. Rev. B}, 95:165102, Apr 2017.

\bibitem{TIbook:2015}
J\'{a}nos~K. Asb\'{o}th, L\'{a}szl\'{o} Oroszl\'{a}ny, and Andr\'{a}s
  P\'{a}lyi.
\newblock {\em A Short Course on Topological Insulators}.
\newblock Springer International Publishing, 2016.

\bibitem{Lodahl:2015}
Immo S{\"o}llner, Sahand Mahmoodian, Sofie~Lindskov Hansen, Leonardo Midolo,
  Alisa Javadi, Gabija Kir{\v{s}}ansk{\.{e}}, Tommaso Pregnolato, Haitham
  El-Ella, Eun~Hye Lee, Jin~Dong Song, S{\o}ren Stobbe, and Peter Lodahl.
\newblock Deterministic photon--emitter coupling in chiral photonic circuits.
\newblock {\em Nature Nanotechnology}, 10(9):775--778, Sep 2015.

\bibitem{Lodahl_Review:2015}
Peter Lodahl, Sahand Mahmoodian, and S\o{}ren Stobbe.
\newblock Interfacing single photons and single quantum dots with photonic
  nanostructures.
\newblock {\em Rev. Mod. Phys.}, 87:347--400, May 2015.

\bibitem{Pozar}
David~M. Pozar.
\newblock {\em Microwave Engineering}.
\newblock Wiley, 4 edition, 2011.

\end{thebibliography}

\end{document}